\pdfoutput=1

\documentclass[preprint,fleqn,5p,numbers,sort&compress]{elsarticle}

\usepackage{graphicx}
\usepackage{amssymb,amsmath}
\usepackage{natbib}
\usepackage{hyperref}
\hypersetup{pdftitle = The title of my PDF, pdfauthor = My name, pdfsubject= The subject, pdfkeywords = keyword1 keyword2 keyword3}
\hypersetup{colorlinks = true, linkcolor = blue, anchorcolor = red, citecolor = blue, filecolor = red, pagecolor = red, urlcolor = blue}

\journal{International Journal of Non-Linear Mechanics}

\newcommand{\norm}[1]{\left\lVert#1\right\rVert}

\begin{document}

\begin{frontmatter}

\title{On the classification of orbits in the three-dimensional Copenhagen problem \\
with oblate primaries}

\author[eez]{Euaggelos E. Zotos\corref{cor1}}
\ead{evzotos@physics.auth.gr}

\author[jn1,jn2]{Jan Nagler}
\ead{jnagler@ethz.ch}

\cortext[cor1]{Corresponding author}

\address[eez]{Department of Physics, School of Science,
Aristotle University of Thessaloniki, GR-541 24, Thessaloniki, Greece}

\address[jn1]{Deep Dynamics Group \& Centre for Human and Machine Intelligence, 
Frankfurt School of Finance \& Management, Frankfurt, Germany}

\address[jn2]{Computational Physics for Engineering Materials,
IfB, ETH Zurich, Wolfgang-Pauli-Strasse 27, CH 8093 Zurich, Switzerland}

\begin{abstract}
The character of motion for the three-dimensional circular restricted three-body problem with oblate primaries is investigated. The orbits of the test particle are classified into four types: non-escaping regular orbits around the primaries, trapped chaotic (or sticky) orbits, escaping orbits that pass over the Lagrange saddle points $L_2$ and $L_3$, and orbits that lead the test particle to collide with one of the primary bodies. We numerically explore the motion of the test particle by presenting color-coded diagrams, where the initial conditions are mapped to the orbit type and studied as a function of the total orbital energy, the initial value of the $z$-coordinate and the oblateness coefficient. The fraction of the collision orbits, measured on the color-coded diagrams, show an algebraic dependence on the oblateness coefficient, which can be derived by simple semi-theoretical arguments.
\end{abstract}

\begin{keyword}
methods: numerical -- celestial mechanics -- spatial restricted three-body problem -- chaos
\end{keyword}
\end{frontmatter}

\section{Introduction}
\label{intro}

A simple model in celestial mechanics is the planar restricted three-body problem (RTBP), that was introduced early on by
Euler. In the model, two primary bodies rotate, around their center of mass in a plane and a test particle is restricted to
this plane and moves only under the gravitational attraction of the primaries \cite{S67}. The chaotic dynamics of the test particle results from its non-integrability and it is a very active field of research, until today (e.g.,
\cite{B03,BvD94,BUF97,H99,H00,H65a,H65b,H69,H97,H01,LBF97,L02,LSNB85,MD94,M89,R01,S67,S78,SM81,SZ93}). Both theoretical and numerical results are presented in the famous papers of Szebehely (e.g., \cite{S67,S78,SM81,LSNB85,SZ93}) and H\'{e}non (e.g., \cite{H65a,H65b,H69,H97,H01}), where the two pioneers devoted major parts to find, describe, and classify periodic orbits. The RTBP became a prototypical model for deterministic chaos (e.g., \cite{R01}). Extensions considered relativistic dynamics (e.g., \cite{B03,MD94}), quantum mechanics (e.g., \cite{BvD94}), and chemical (e.g, \cite{BUF97,LBF97}), or even astrophysical (e.g., \cite{H99,H00,M89}). The RTBP has been used for investigating the stability of several types of solar systems (e.g., \cite{L02}), chaos-assisted asteroid capture (e.g., \cite{ABWF03}), as well as for modelling the orbital dynamics of a pair of black holes, orbited by a sun (e.g., \cite{Q96}) -- to mention only a few applications.

Usually, collision orbits can be considered as some kind of ``leakage" of the phase space (e.g., \cite{APT13,BBS09,C90,CK92,CKK93,EP14,KSCD99,NKL07,NH01,STN02,Z14}). In the classical version of the restricted three-body problem the shape of the two primary bodies is assumed, in most of the cases, to be spherically symmetric. However, in nature, and particularly in space, there are many celestial bodies, such as Saturn and Jupiter, which have in fact oblate shape (e.g., \cite{BPC99}) or even a triaxial one. Moreover, completely irregular shapes also exist. Such celestial bodies with irregular shapes are mainly asteroids and meteoroids (e.g., \cite{MWF87,NC08}). The main effect on a celestial body, which is directly related to the oblateness and triaxiality, is a perturbation which leads to deviation from the classical two-body motion. The influence of the oblateness coefficient, when the more massive primary body is an oblate spheroid, has been studied in a series of works (e.g., \cite{BS12,KMP05,KDP06,KPP08,MPP96,MRVK00,PPK12,S81,S87,S89,S90,SSR79,SSR86,SL12,SL13,SRS88,SRS97}).

By today it is a well known fact that almost all the planets of our solar system are in fact oblate spheroidals. Such deviations from spherical symmetry have been taken into consideration in many recent studies. For instance for the Saturn-Dione-satellite and Saturn-Tethys-satellite systems it was proved in \cite{OV03} that the convergence between the theoretical data and the corresponding ones from observations is much more better, when the oblateness of Saturn is included in the equations. In the same vein, in \cite{SY08} the oblateness of Neptune was taken into consideration for modeling the motion of the spacecraft in the Neptune-Triton system.

In this work, we numerically investigate two aspects of the RTBP. First, the three dimensional version of this model is studied, where the test particle is not restricted to the plane but it can move freely in three dimensions. This extension was also investigated in \cite{Z16}. Second, being the main novelty, we determine the impact of oblateness of the primary bodies on escape orbits, bounded orbits, and collision orbits. Our study complements previous works, where bounded, escape and collision motion was studied in a series of papers for the 2-dof RTBP (e.g. \cite{Z15a,Z15b,Z15c}).

The present paper is structured as follows: the basic properties of the dynamical system are presented in Section \ref{mod}. All the computational methodology which will be used is described in detail in Section \ref{cometh}. The following Sections contains all the numerical results, regarding the orbit classification in the 3-dof RTBP. Our paper ends with Section \ref{conc}, where the concluding remarks are emphasized.

\section{Description of the dynamical system}
\label{mod}

\begin{figure}[!t]
\centering
\resizebox{\hsize}{!}{\includegraphics{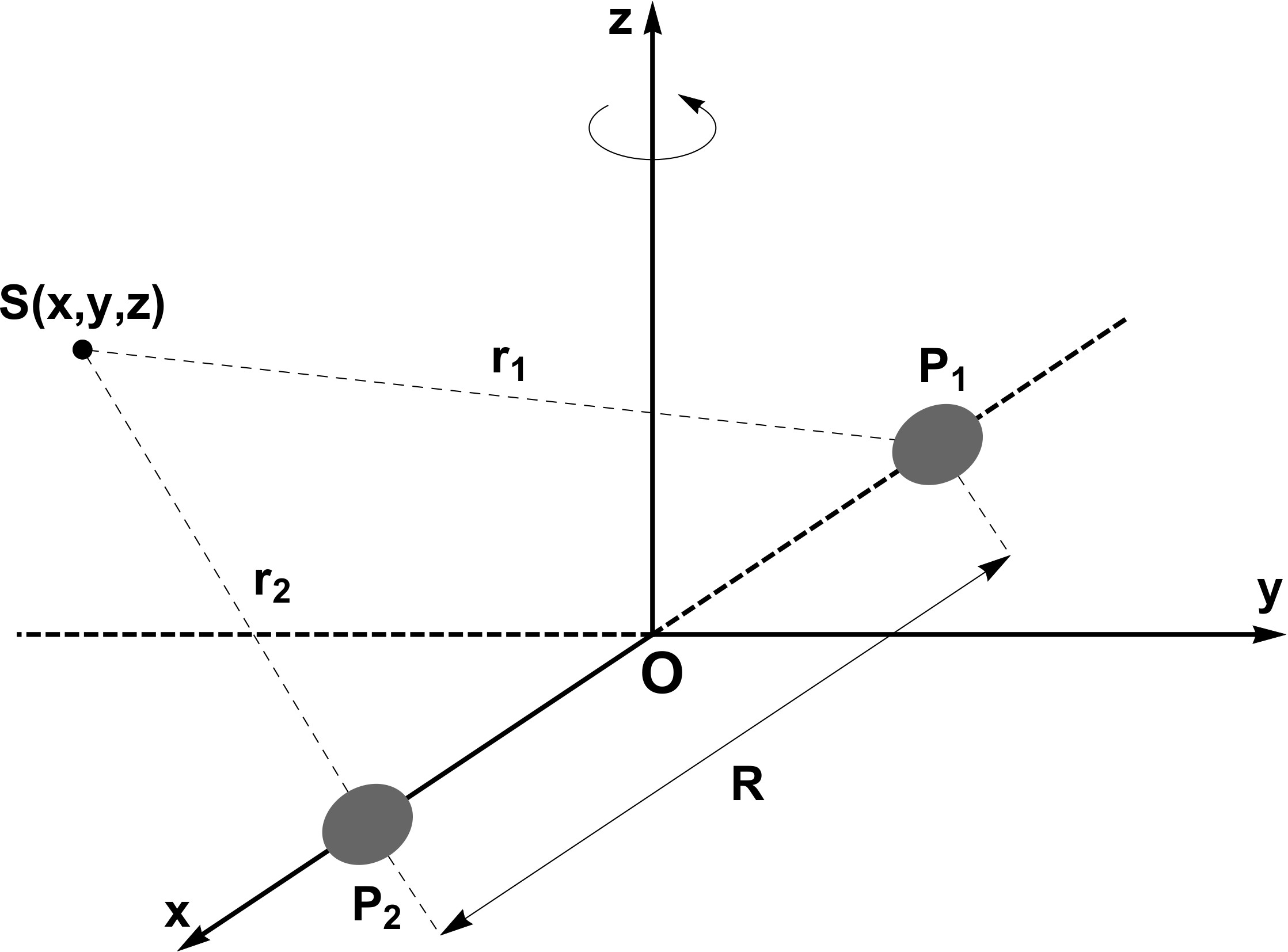}}
\caption{A schematic depicting the space configuration of the circular restricted three-body problem, when the primary bodies are oblate spheroids.}
\label{sch}
\end{figure}

The dynamical system consists of two primary bodies, $P_1$ and $P_2$, which perform circular Keplerian orbits around their common mass center \cite{S67} (see Fig. \ref{sch}). The third body moves under the combined gravitational attraction of the two primaries. Considering that the mass of the third body $m$ is considerable smaller, with respect to the masses of the two primary bodies, we may assume the motion of the primaries is not perturbed, in any way, by the test particle.

In our system the gravitational constant $G$, the distance $R$ between the primaries and the sum of their masses are equal to unity. The dimensionless masses of the primary bodies are $m_1 = 1 - \mu$ and $m_2 = \mu$, where of course $\mu = m_2/(m_1 + m_2) \leq 1/2$ is the mass parameter. Furthermore, the centers of both primary bodies lie on the $x$-axis, at $(x_1, 0, 0)$ and $(x_2, 0, 0)$, where $x_1 = - \mu$ and $x_2 = 1 - \mu$. In the present study we shall consider the Copenhagen case, that is the case where the primaries have equal masses $m_1 = m_2$ and therefore $\mu = 1/2$. We consider a dimensionless, barycentric, rotating system of coordinates $Oxyz$, in which the $Ox$ axis always contains the two primary bodies, while the center of mass coincides with the origin $(0,0,0)$.

According to \cite{SSR75,OV03,DM06,AS06} the time-independent effective potential function of the RTBP with oblate primaries is
\begin{align}
\Omega(x,y,z) &= \sum_{i=1}^{2} \frac{m_i}{r_i}\left(1 + \frac{A_i}{2r_i^2} - \frac{3A_i z^2}{2r_i^4}\right) \nonumber\\
& + \frac{n^2}{2} \left(x^2 + y^2 \right) + \frac{\mu\left(1 - \mu \right)}{2},
\label{pot}
\end{align}
where
\begin{align}
r_1 &= \sqrt{\left(x + \mu \right)^2 + y^2 + z^2}, \nonumber\\
r_2 &= \sqrt{\left(x + \mu -1 \right)^2 + y^2 + z^2},
\label{dist}
\end{align}
are the distances of the third body from the respective primaries. Moreover, $A_i$, $i = 1, 2$ are the oblateness coefficients, while $n$ is the angular velocity which is defined as
\begin{equation}
n = \sqrt{1 + 3\left(A_1 + A_2 \right)/2}.
\label{vel}
\end{equation}
In our study we shall consider the case of relatively low values of oblateness coefficient, where $A_1 = A_2 = 10^{-4}$. This choice can be justified if we take into consideration that the values of the oblateness in our Solar System are very low (see e.g., Table 1 in \cite{SSR76}).

The equations describing the motion of the test particle, in the corotating frame of reference, read
\begin{align}
\ddot{x} &= \frac{\partial \Omega}{\partial x} + 2 n \dot{y}, \nonumber\\
\ddot{y} &= \frac{\partial \Omega}{\partial y} - 2 n \dot{x}, \nonumber\\
\ddot{z} &= \frac{\partial \Omega}{\partial z}.
\label{eqmot}
\end{align}

\begin{figure*}[!t]
\centering
\resizebox{\hsize}{!}{\includegraphics{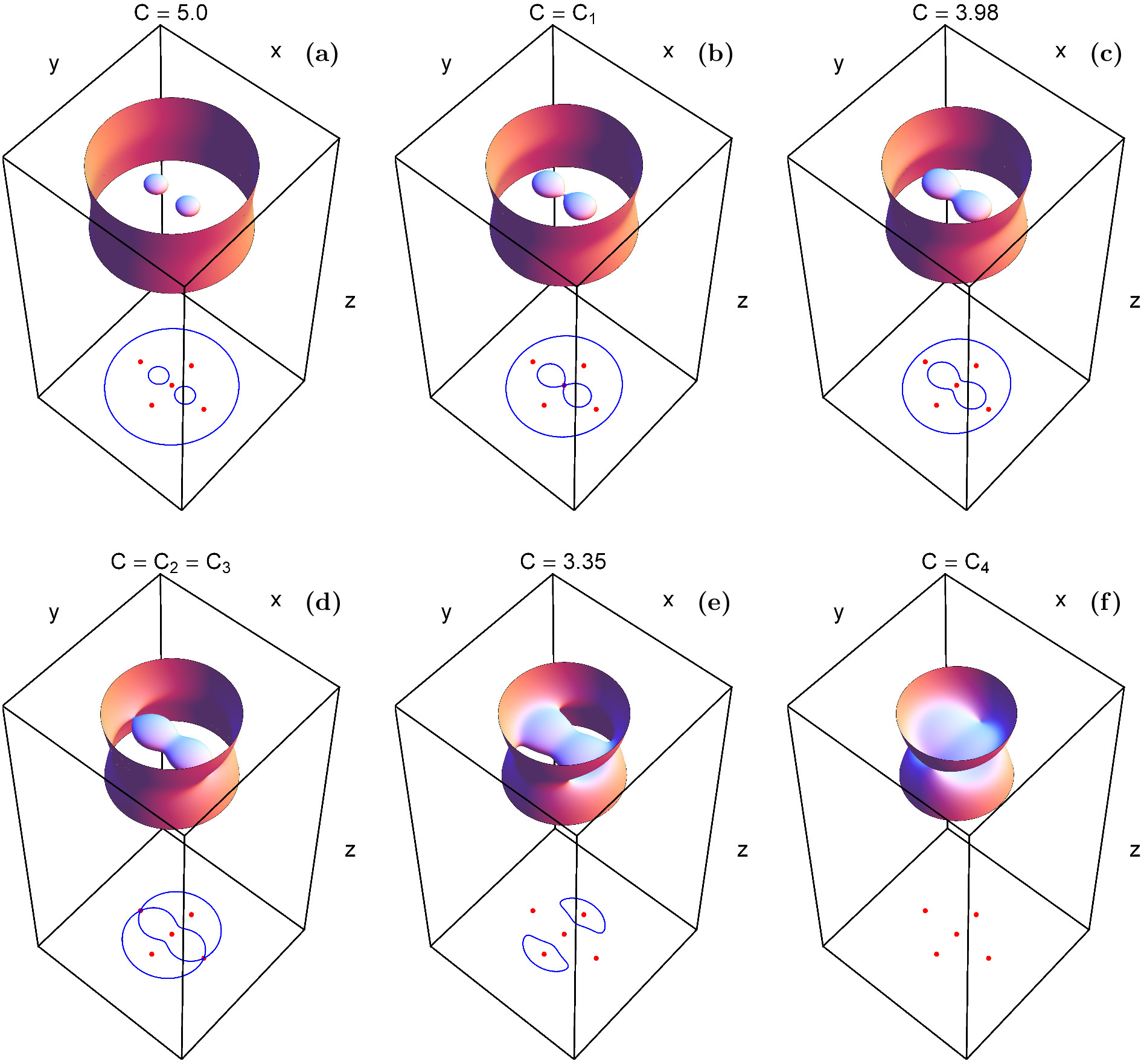}}
\caption{Evolution of the structure of three-dimensional zero velocity surfaces, as a function of the Jacobi constant $C$. The projections of these surfaces onto the configuration $(x,y)$ plane are depicted, at the bottom of the bounding box, as blue solid lines. The projected positions of the five coplanar equilibrium points $L_i$, $i=1,...,5$ are pinpointed by red dots. (Color figure online).}
\label{zvs}
\end{figure*}

The Jacobi integral of motion is described by the Hamiltonian
\begin{equation}
J(x,y,z,\dot{x},\dot{y},\dot{z}) = 2\Omega(x,y,z) - \left(\dot{x}^2 + \dot{y}^2 + \dot{z}^2 \right) = C,
\label{ham}
\end{equation}
where $\dot{x}$, $\dot{y}$, and $\dot{z}$ are the velocities, while $C$ is the conserved value of the Hamiltonian.

In the classical RTBP (that is when $A_1 = A_2 = 0$) there are five equilibrium points, which are also known as Lagrange points. All five equilibrium points are located on the configuration $(x,y)$ plane with $z = 0$. The central point $L_1$ is located between the two primaries, $L_2$ is located at the right side of primary $P_2$ (with $x > 0$), while $L_3$ is located at the left side of primary $P_1$ (with $x < 0$). In addition $L_4$ has $y > 0$, while the libration point $L_5$ has $y < 0$. In our case, where $A_1 = A_2 = 10^{-4}$ the positions of the five equilibrium points are the following:
\begin{itemize}
  \item Equilibrium point $L_1$: $(x,y) = (0,0)$.
  \item Equilibrium point $L_2$: $(x,y) = (1.19839762,0)$.
  \item Equilibrium point $L_3$: $(x,y) = (-1.19839762,0)$.
  \item Equilibrium point $L_4$: $(x,y) = (0,0.86596768)$.
  \item Equilibrium point $L_5$: $(x,y) = (0,-0.86596768)$.
\end{itemize}

The values of the Jacobi constant at the Lagrange points $L_i, i = 1, ..., 5$ are denoted by $C_i$. For the Copenhagen problem with oblateness we have: $C_1 = 4.2508$, $C_2 = C_3 = 3.70738405$, and $C_4 = C_5$ = 3.00032499.

The relation $2\Omega(x,y,z) = C$ defines a three-dimensional surface, which is known as the zero velocity surface (ZVS) (see e.g., \cite{DL13}). The ZVS allows us to determine the energetically possible orbits of the test particle, while its projection on the configuration
$(x,y)$ plane is called Hill's regions. In the same vein, the  boundaries of the Hill's regions are called zero velocity curves (ZVCs), since they are the locus where the kinetic energy is zero. The geometric shape of the ZVSs, and of course of the corresponding ZVCs strongly depends on the Jacobi constant. More precisely, four distinct energy regions exist regarding the configurations of the Hill's regions:
\begin{itemize}
  \item Energy region I: $C > C_1$: All communication channels are closed, so the test particle moves either very close or far
      away from the primaries.
  \item Energy region II: $C_1 > C > C_2$: Only the channel around $L_1$ is open and therefore orbits are allowed to communicate between the primary bodies.
  \item Energy region III: $C_2 > C > C_4$: Both channels around $L_2$ and $L_3$ are open, which implies that the test particle can freely escape, through two directions.
  \item Energy region IV: $C < C_4$: The energetically forbidden regions are confined, while motion over the entire $(x,y)$ plane is possible.
\end{itemize}
The evolution of the geometry of the ZVSs, for several values of the Jacobi constant, is presented in Fig. \ref{zvs}.

\section{Computational methodology}
\label{cometh}

For revealing the orbital properties of the Hamiltonian system we have to determine the nature of several sets of initial conditions of orbits. In 3-dof systems the dynamics along the direction of the $z$ axis is very important. On this basis, we choose to classify initial conditions on the $(x,z)$ plane. More precisely, for several values of the Jacobi constant $C$, we create uniform grids of $1024 \times 1024$ initial conditions $(x_0,z_0)$, inside a square area with $-4 \leq x, z \leq 4$. For all orbits $y_0 = \dot{x_0} = \dot{z_0} = 0$, while the initial value of the velocity $\dot{y}$ is calculated as follows
\begin{equation}
\dot{y_0}(x_0,y_0,z_0;C) = - \sqrt{2 \Omega(x_0, y_0 = 0, z = z_0) - C}.
\label{ini}
\end{equation}
This particular choice of the initial conditions has a significant advantage, which will be presented and explained in the
following Section.

The nature of the initial conditions of the orbits can be categorized as follows:
\begin{enumerate}
  \item Bounded orbits.
  \item Escaping orbits.
  \item Orbits that collide with one of the primaries.
\end{enumerate}
Furthermore, bounded orbits will be further classified into three sub-categories:
\begin{enumerate}
  \item Non-escaping regular orbits.
  \item Trapped sticky orbits.
  \item Trapped chaotic orbits.
\end{enumerate}

It should be noted, that there is also the type of orbits which pass over the inner equilibrium point $L_1$. However, we will not include this type of orbits in our study. This choice is justified because our classification reveals the final state of the orbits. On the other hand, orbits that pass over $L_1$ can later collide with one of the primaries, or escape from the system through $L_2$ or $L_3$, or even display trapped chaotic motion around the primaries. Therefore, passing through $L_1$ is only a first stage, while in this work we are interesting about the final state of the orbits. Moreover, we follow the same orbit classification which has been adopted in previous related papers (e.g., \cite{N04,N05,Z15a}).

For determining the regular or chaotic character of the orbits we use the Smaller Alignment Index (SALI) \cite{S01}, which is  defined as
\begin{equation}
\rm SALI(t) \equiv min(d_-, d_+),
\label{sali}
\end{equation}
where
\begin{align}
d_- &\equiv \norm{ \frac{{\vec{w_1}}(t)}{\| {\vec{w_1}}(t) \|} - \frac{{\vec{w_2}}(t)}{\| {\vec{w_2}}(t) \|} }, \nonumber\\
d_+ &\equiv \norm{ \frac{{\vec{w_1}}(t)}{\| {\vec{w_1}}(t) \|} + \frac{{\vec{w_2}}(t)}{\| {\vec{w_2}}(t) \|} },
\label{align}
\end{align}
are the alignments indices, while ${\vec{w_1}}(t)$ and ${\vec{w_2}}(t)$, are two deviation vectors which initially are orthonormal and point in two random directions. The variational equations needed for the computation of the two deviation vectors are given by
\begin{align}
\dot{(\delta x)} &= \delta \dot{x}, \nonumber\\
\dot{(\delta y)} &= \delta \dot{y}, \nonumber\\
\dot{(\delta z)} &= \delta \dot{z}, \nonumber\\
(\dot{\delta \dot{x}}) &= \frac{\partial^2 \Omega}{\partial x^2}\delta x + \frac{\partial^2 \Omega}{\partial x \partial
y}\delta y + \frac{\partial^2 \Omega}{\partial x \partial z}\delta z + 2 n \delta \dot{y}, \nonumber \\
(\dot{\delta \dot{y}}) &= \frac{\partial^2 \Omega}{\partial y \partial x}\delta x + \frac{\partial^2 \Omega}{\partial
y^2}\delta y + \frac{\partial^2 \Omega}{\partial y \partial z}\delta z - 2 n \delta \dot{x}, \nonumber\\
(\dot{\delta \dot{z}}) &= \frac{\partial^2 \Omega}{\partial z \partial x}\delta x + \frac{\partial^2 \Omega}{\partial z
\partial y}\delta y + \frac{\partial^2 \Omega}{\partial z^2}\delta z.
\label{variac}
\end{align}

For distinguishing between bounded and unbounded (escaping) motion we consider a limiting sphere with center at the origin
$(0,0,0)$ and radius $R_{d}$. Then the case of bounded motion is present if the test particle stays confined, for integration time $t_{\rm max}$, inside the limiting sphere. On the other hand, escaping motion occurs if the test particle intersects the limiting sphere with velocity pointing outwards. For our calculations we choose $t_{\rm max} = 10^4$ dimensionless time units and $R_d = 10$. Moreover, a collision occurs if the test particle crosses the sphere with radius $R_{\rm col}$ around the primaries, where $R_{\rm col} = 10^{-4}$. For both the parameters $R_d$ and $R_{\rm col}$ we adopted the values also used in \cite{N04,N05,Z15a,Z15c}.

For the numerical integration of the equations of motion (\ref{eqmot}) and the corresponding variational equations (\ref{variac}) we used a double precision Bulirsch-Stoer \verb!FORTRAN 77! algorithm \cite{PTVF92}. In all our calculations the numerical error, related to the conservation of the Jacobi integral (\ref{ham}) was generally smaller than $10^{-12}$, while in most of the cases it was smaller than $10^{-14}$. All the graphical illustration has been created using the latest version 11.2 of the software Mathematica$^{\circledR}$ \cite{W03}.

\section{Orbital dynamics}
\label{numres}

In this Section we shall present color-coded diagrams (CCDs) thus following the methods introduced and used in \cite{N04} and \cite{N05}. In these diagrams, each pixel corresponds to a specific initial condition and it is colored according to the classification of the particular orbit. Throughout the section we will use several technical terms for describing the orbital properties of the system. The most important terminology is explained in \ref{apex}, where we also describe how one can interpret the different aspects of the color-coded diagrams.

\subsection{Energy case I $(C \geq C_1)$}
\label{ss1}

\begin{figure*}[!t]
\centering
\resizebox{\hsize}{!}{\includegraphics{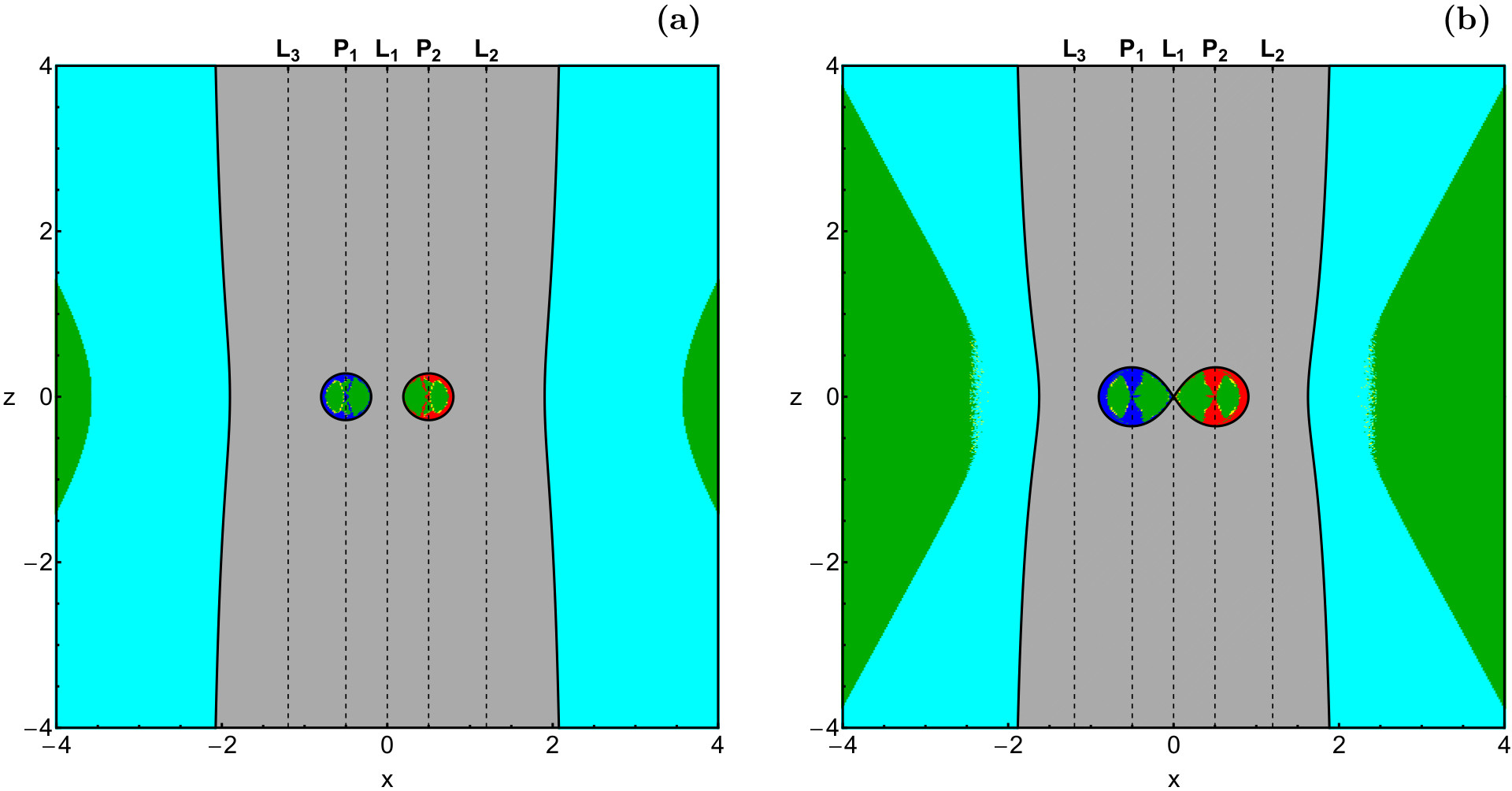}}
\caption{The orbital structure of the configuration $(x,z)$ plane, for the energy case I, when (a-left): $C = 5.0$ and
(b-right): $C = C_1$. The color code is the following: bounded regular orbits (green), trapped chaotic orbits (yellow),
collision orbits to primary 1 (blue), collision orbits to primary 2 (red), and escaping orbits (cyan). The energetically forbidden
regions are shown in gray. (Color figure online.)}
\label{HR1}
\end{figure*}

\begin{figure*}[!t]
\centering
\resizebox{\hsize}{!}{\includegraphics{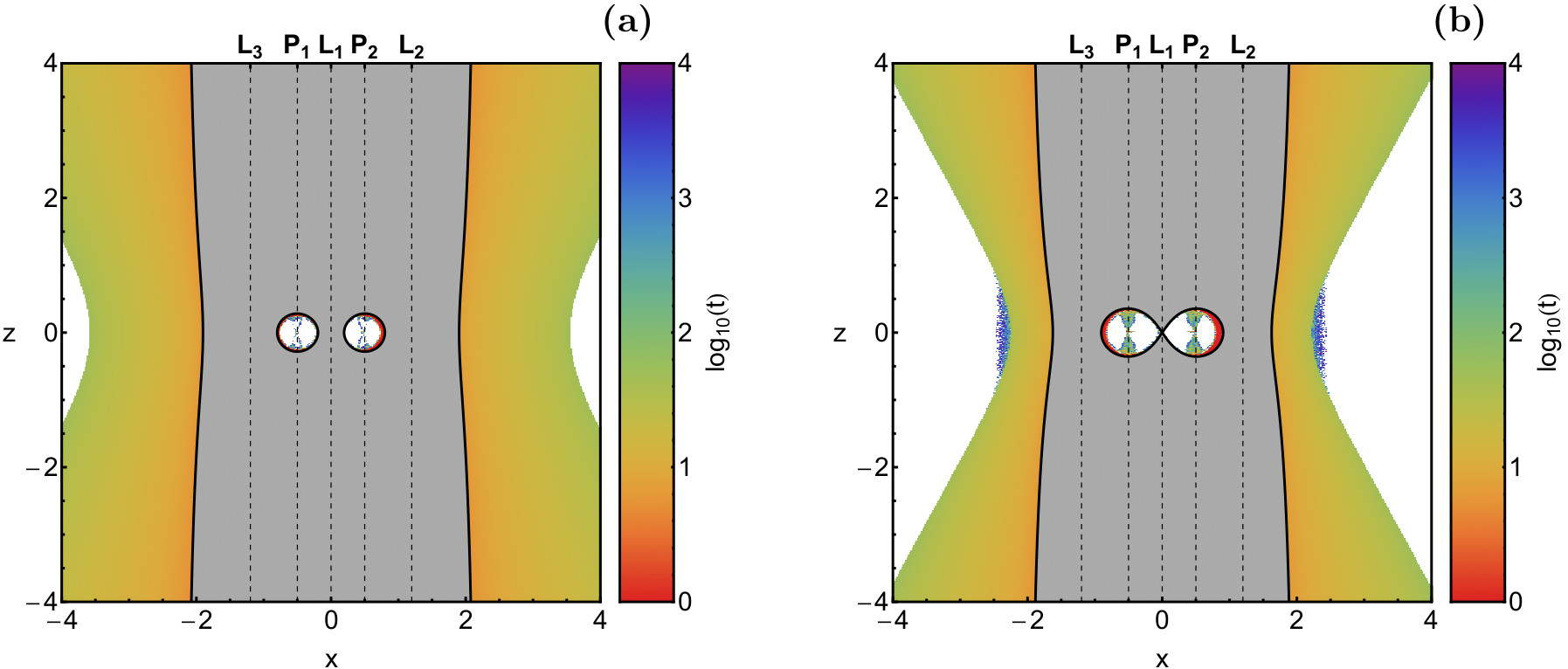}}
\caption{The corresponding distribution of the escape and collision time of the orbits for the values of the Jacobi constant of Fig. \ref{HR1}(a-b). The darker the color, the larger the escape/collision time. Initial conditions of bounded regular orbits and trapped chaotic orbits are shown in white. (Color figure online.)}
\label{HR1t}
\end{figure*}

\begin{figure*}[!t]
\centering
\resizebox{\hsize}{!}{\includegraphics{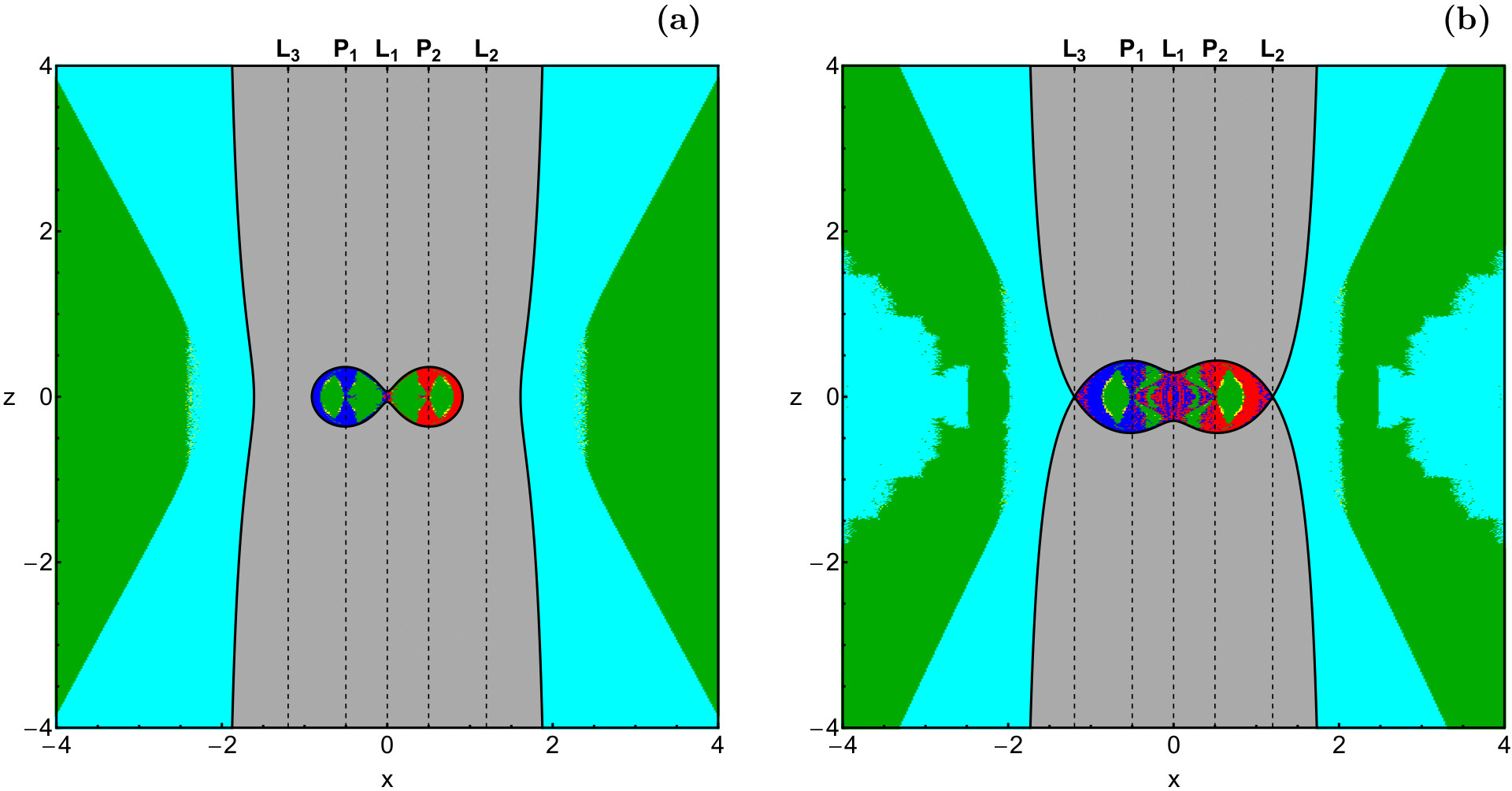}}
\caption{The orbital structure of the the configuration $(x,z)$ plane, for the energy case II, when (a-left): $C = 4.22$ and
(b-right): $C = C_2$. The color code is the same as in Fig. \ref{HR1}. (Color figure online.)}
\label{HR2}
\end{figure*}

\begin{figure*}[!t]
\centering
\resizebox{\hsize}{!}{\includegraphics{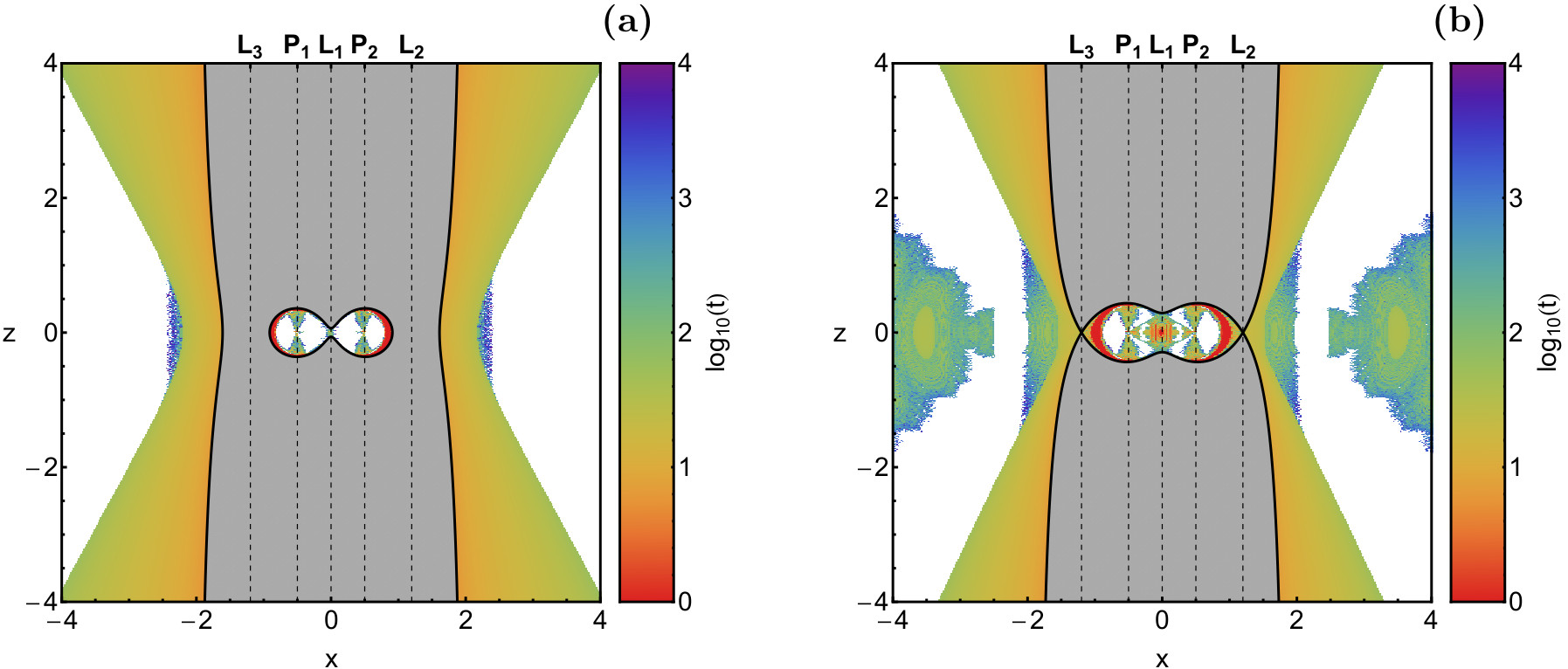}}
\caption{The corresponding distribution of the escape and collision time of the orbits for the values of the Jacobi constant of Fig. \ref{HR2}(a-b). (Color figure online.)}
\label{HR2t}
\end{figure*}

\begin{figure*}[!t]
\centering
\resizebox{\hsize}{!}{\includegraphics{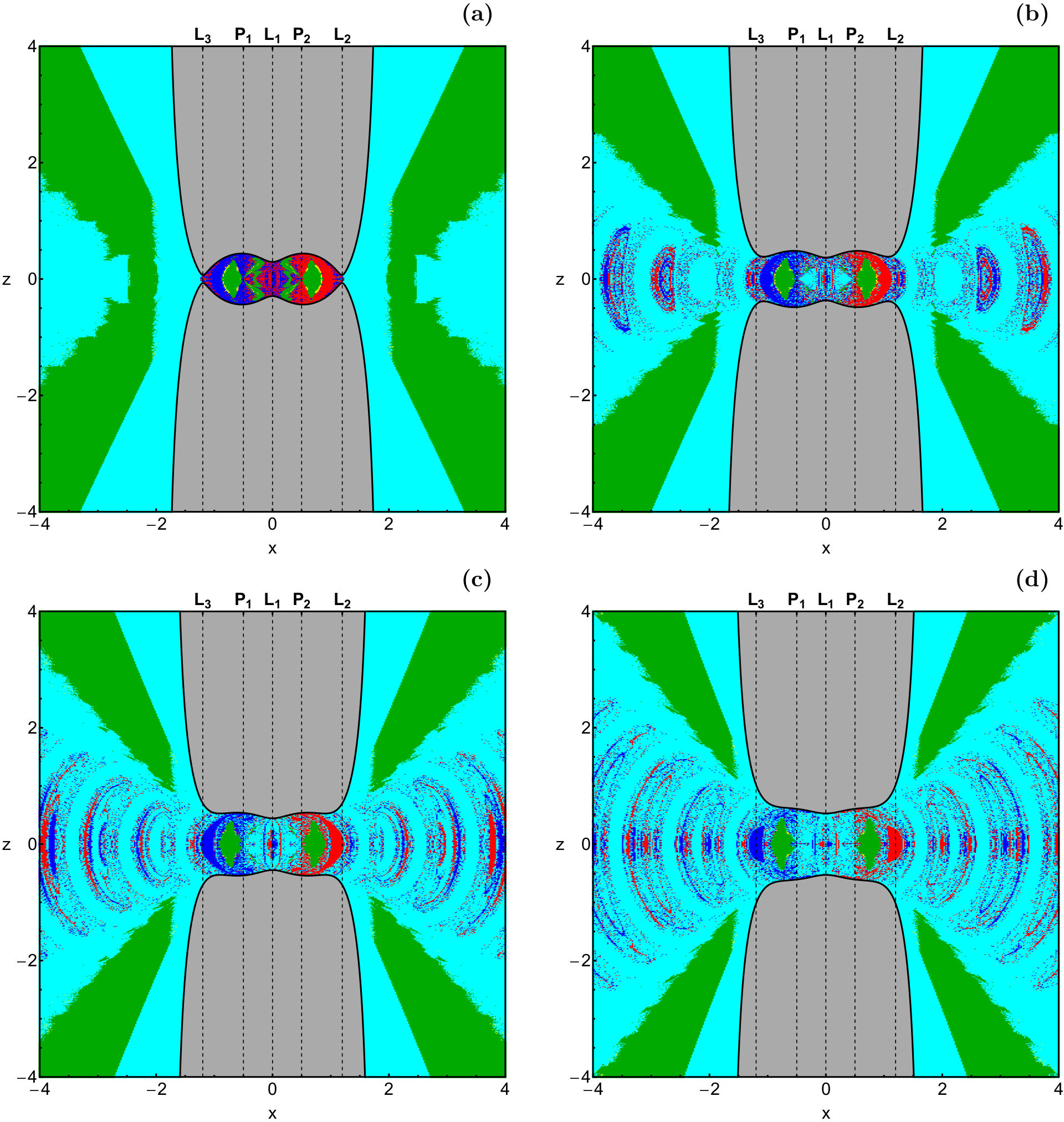}}
\caption{The orbital structure of the configuration $(x,z)$ plane, for the energy case III, when (a-upper left): $C = 3.70$;
(b-upper right): $C = 3.47$; (c-lower left): $C = 3.24$; and (d-lower right): $C = C_4$. The color code is the same as in Fig. \ref{HR1}. (Color figure online.)}
\label{HR3}
\end{figure*}

\begin{figure*}[!t]
\centering
\resizebox{\hsize}{!}{\includegraphics{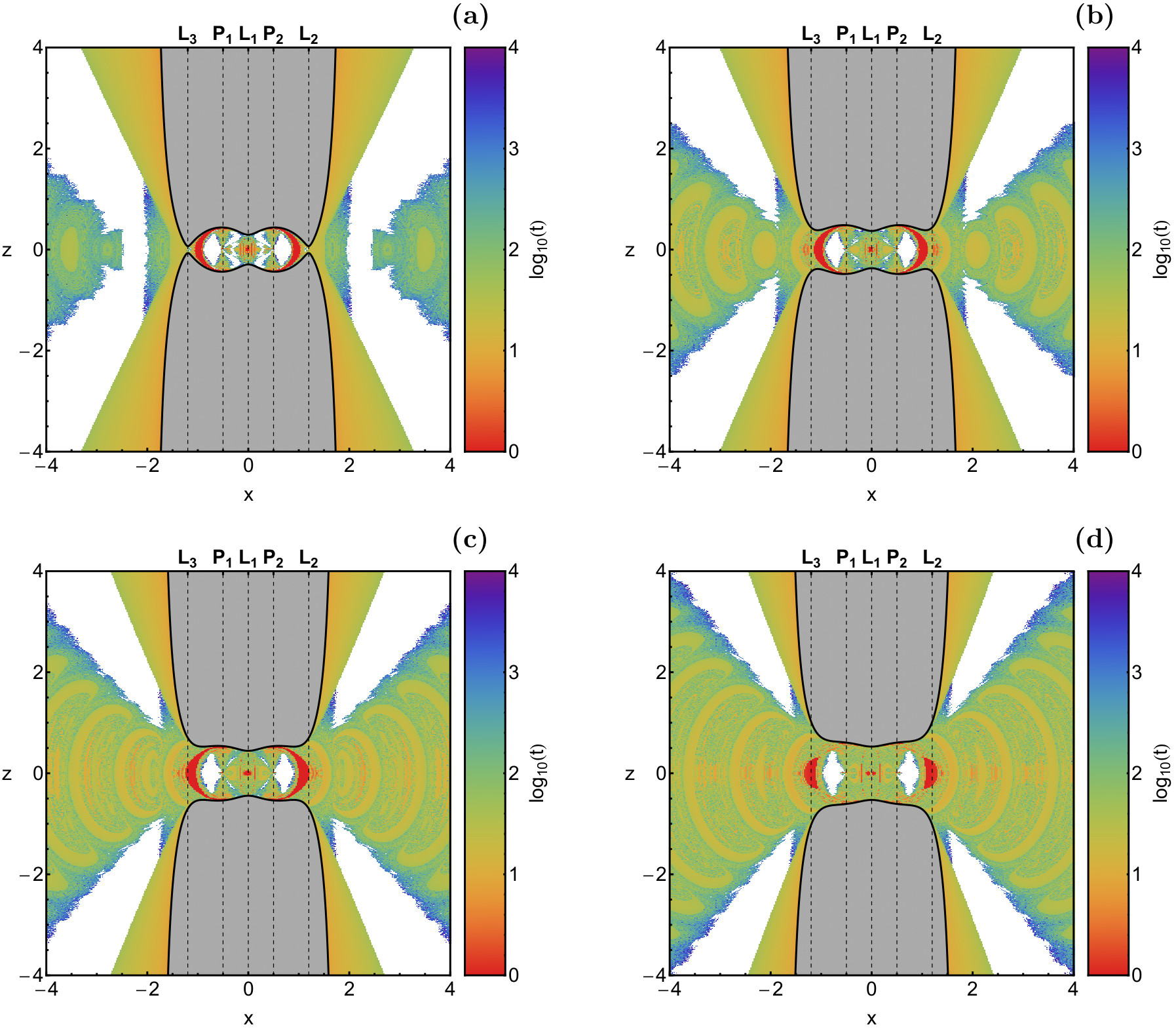}}
\caption{The corresponding distribution of the escape and collision time of the orbits for the values of the Jacobi constant of Fig. \ref{HR3}(a-d). (Color figure online.)}
\label{HR3t}
\end{figure*}

\begin{figure*}[!t]
\centering
\resizebox{\hsize}{!}{\includegraphics{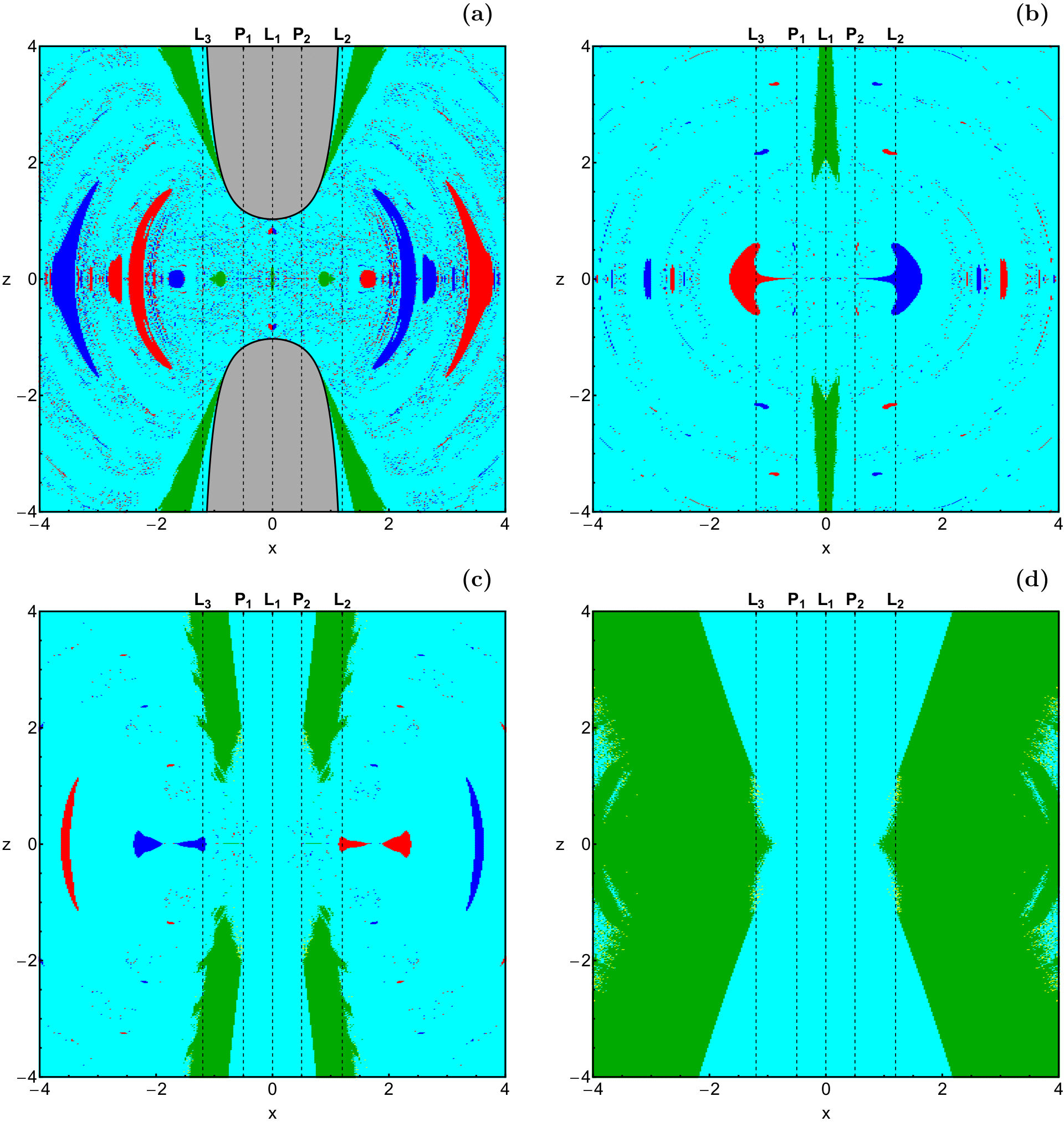}}
\caption{The orbital structure of the configuration $(x,z)$ plane, for the energy case IV, when (a-upper left): $C = 2$;
(b-upper right): $C = 0.5$; (c-lower left): $C = -0.5$; and (d-lower right): $C = -2$. The color code is the same as in Fig.
\ref{HR1}. (Color figure online.)}
\label{HR4}
\end{figure*}

\begin{figure*}[!t]
\centering
\resizebox{\hsize}{!}{\includegraphics{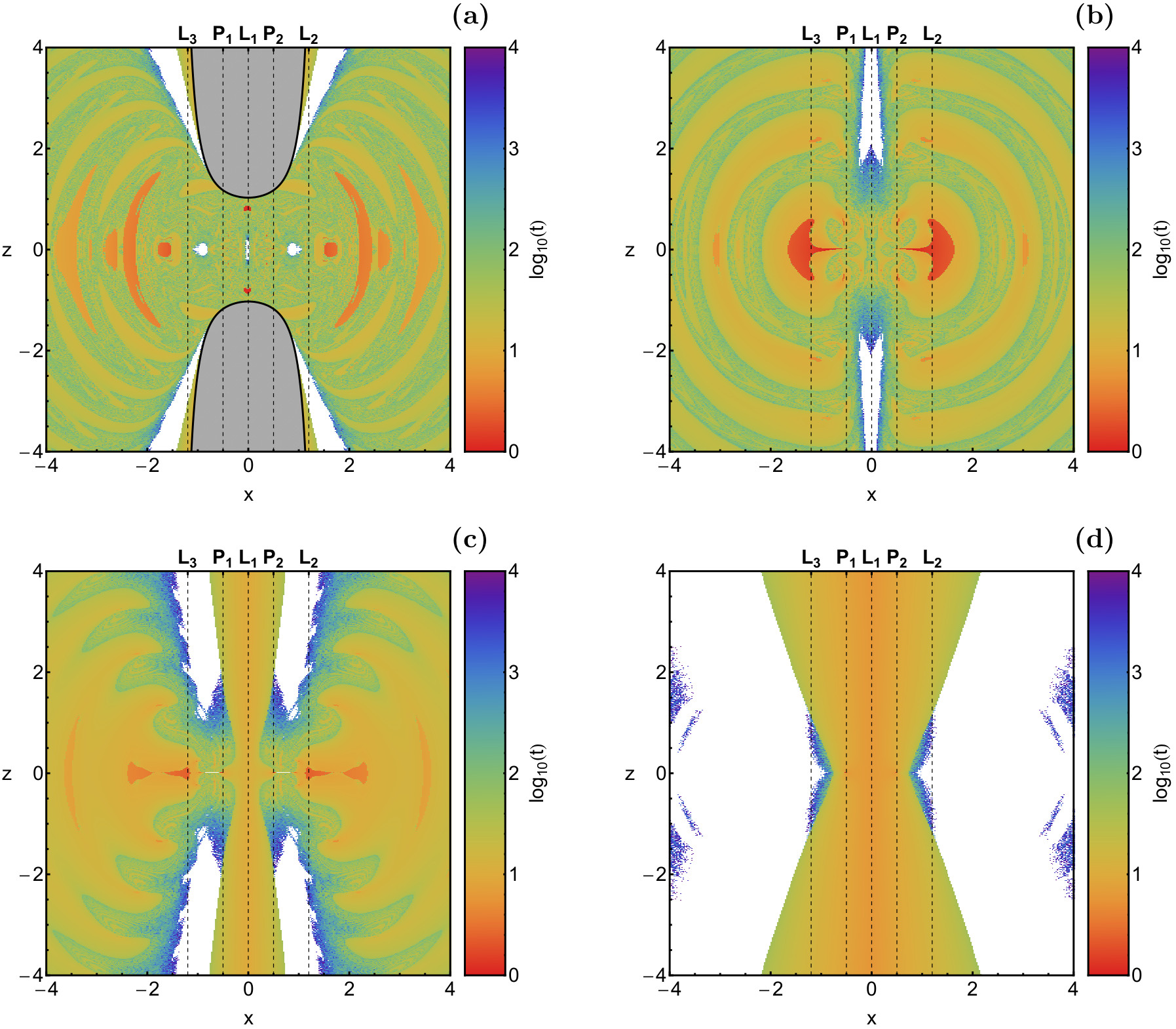}}
\caption{The corresponding distribution of the escape and collision time of the orbits for the values of the Jacobi constant of Fig. \ref{HR3}(a-d). (Color figure online.)}
\label{HR4t}
\end{figure*}

\begin{figure*}[!t]
\centering
\resizebox{0.70\hsize}{!}{\includegraphics{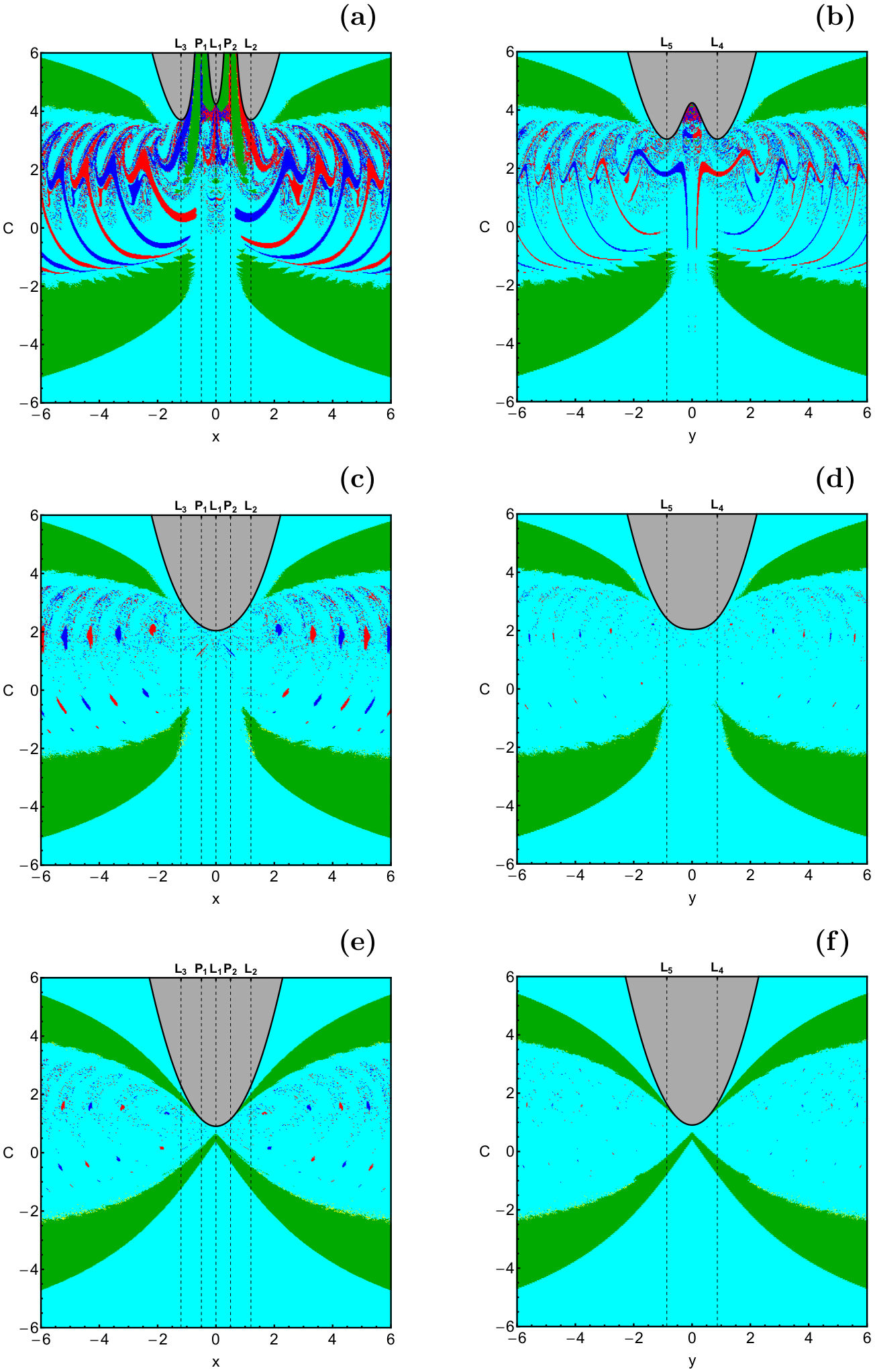}}
\caption{Orbital structure of the (first column): $(x,C)$ plane and (second column): $(y,C)$ plane, when (first row): $z_0 = 0.01$; (second row): $z_0 = 1$; and (third row): $z_0 = 3$. The color code is the same as in Fig. \ref{HR1}. (Color figure online.)}
\label{xyC}
\end{figure*}

\begin{figure*}[!t]
\centering
\resizebox{0.80\hsize}{!}{\includegraphics{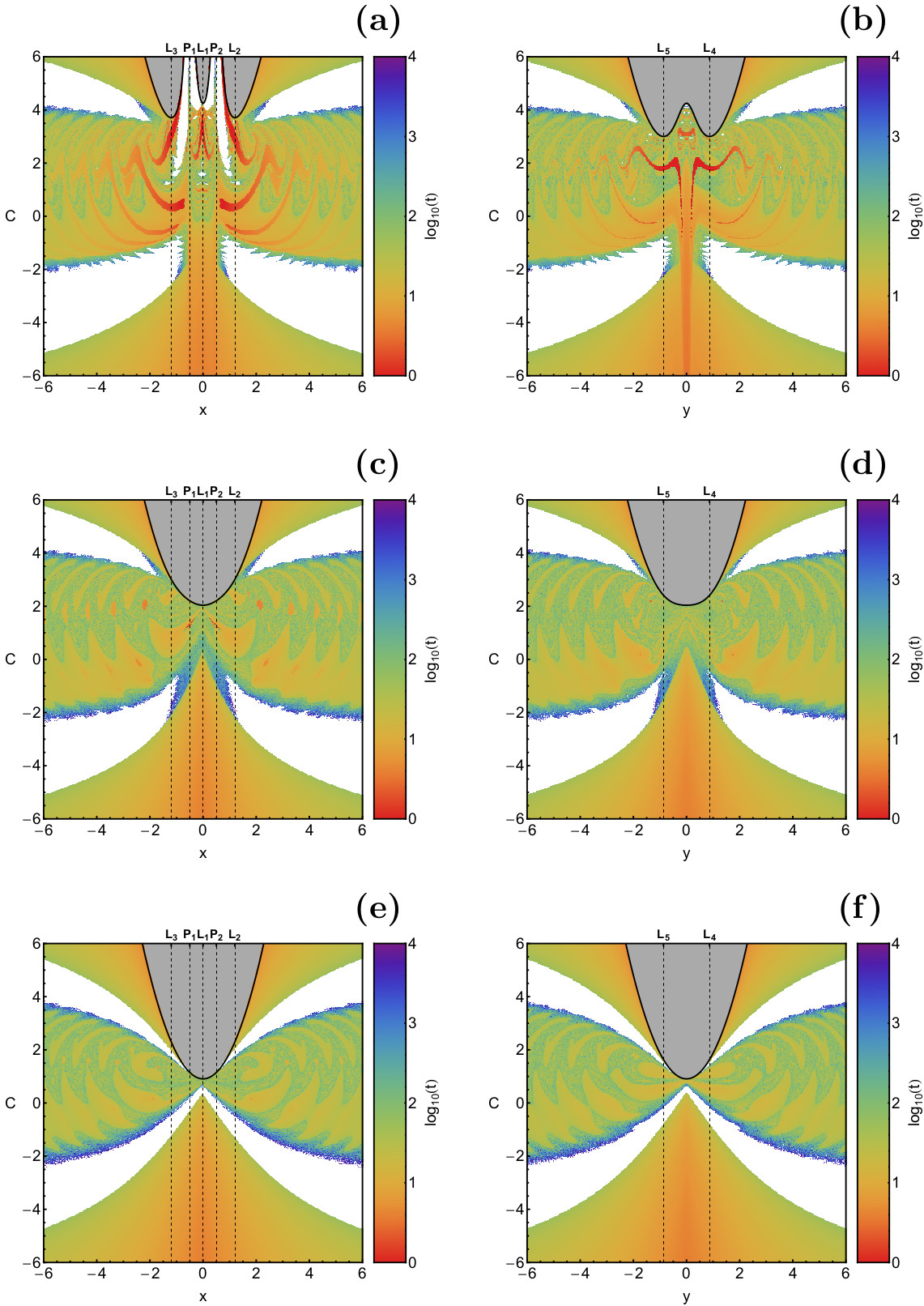}}
\caption{The corresponding distribution of the escape and collision time of the orbits for the panels of Fig. \ref{xyC}. (Color figure online.)}
\label{xyCt}
\end{figure*}

\begin{figure*}[!t]
\centering
\resizebox{0.70\hsize}{!}{\includegraphics{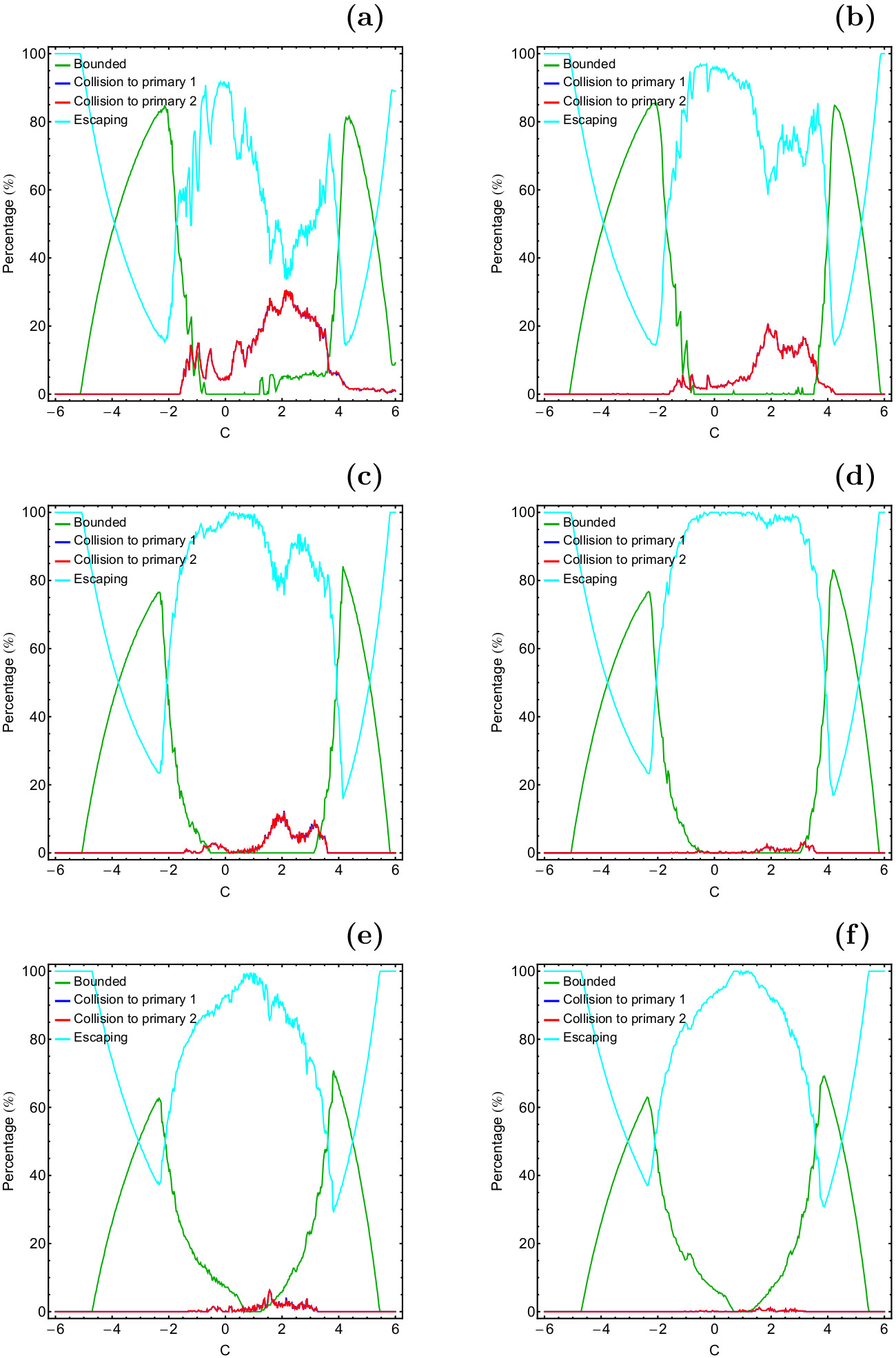}}
\caption{Evolution of the percentages of escaping, bounded and collision orbits on the $(x,C)$ and $(y,C)$ planes, for the panels of Fig. \ref{xyC}, as a function of the value of the Jacobi constant $C$. Note that because the amount of collision orbits to primary $P_1$ completely coincides with the amount of orbits that lead to collision with primary $P_2$ the corresponding percentages are the same and therefore the two curves on the diagram (blue and red) are overlapping each other. (Color figure online.)}
\label{pxyC}
\end{figure*}

\begin{figure*}[!t]
\centering
\resizebox{0.70\hsize}{!}{\includegraphics{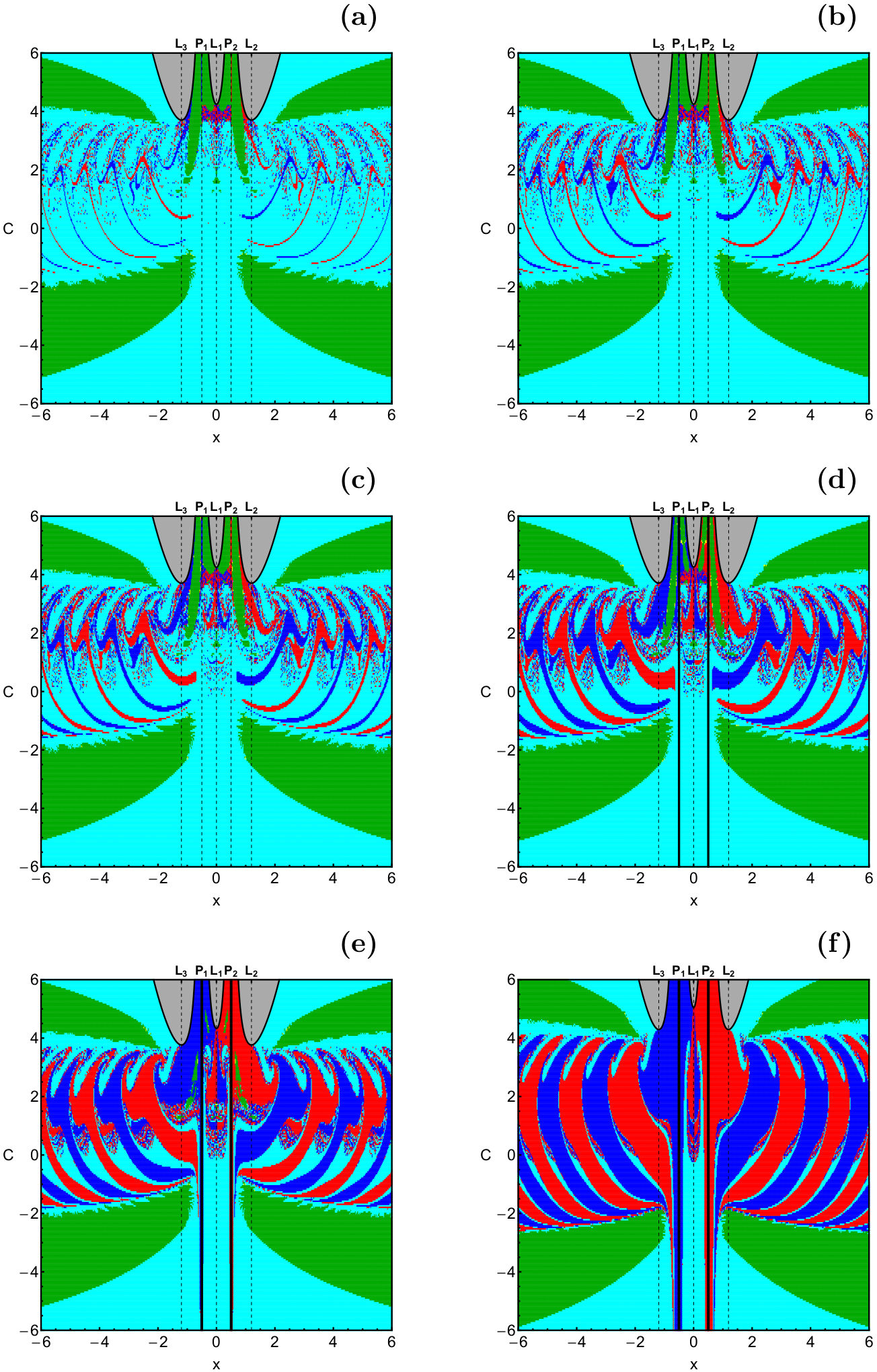}}
\caption{Orbital structure of the $(x,C)$ plane when (a): $A = 10^{-6}$; (b): $A = 10^{-5}$; (c): $A = 10^{-4}$; (d): $A = 10^{-3}$; (e): $A = 10^{-2}$; (f): $A = 10^{-1}$. The color code is the same as in Fig. \ref{HR1}. (Color figure online.)}
\label{xCA}
\end{figure*}

\begin{figure*}[!t]
\centering
\resizebox{0.80\hsize}{!}{\includegraphics{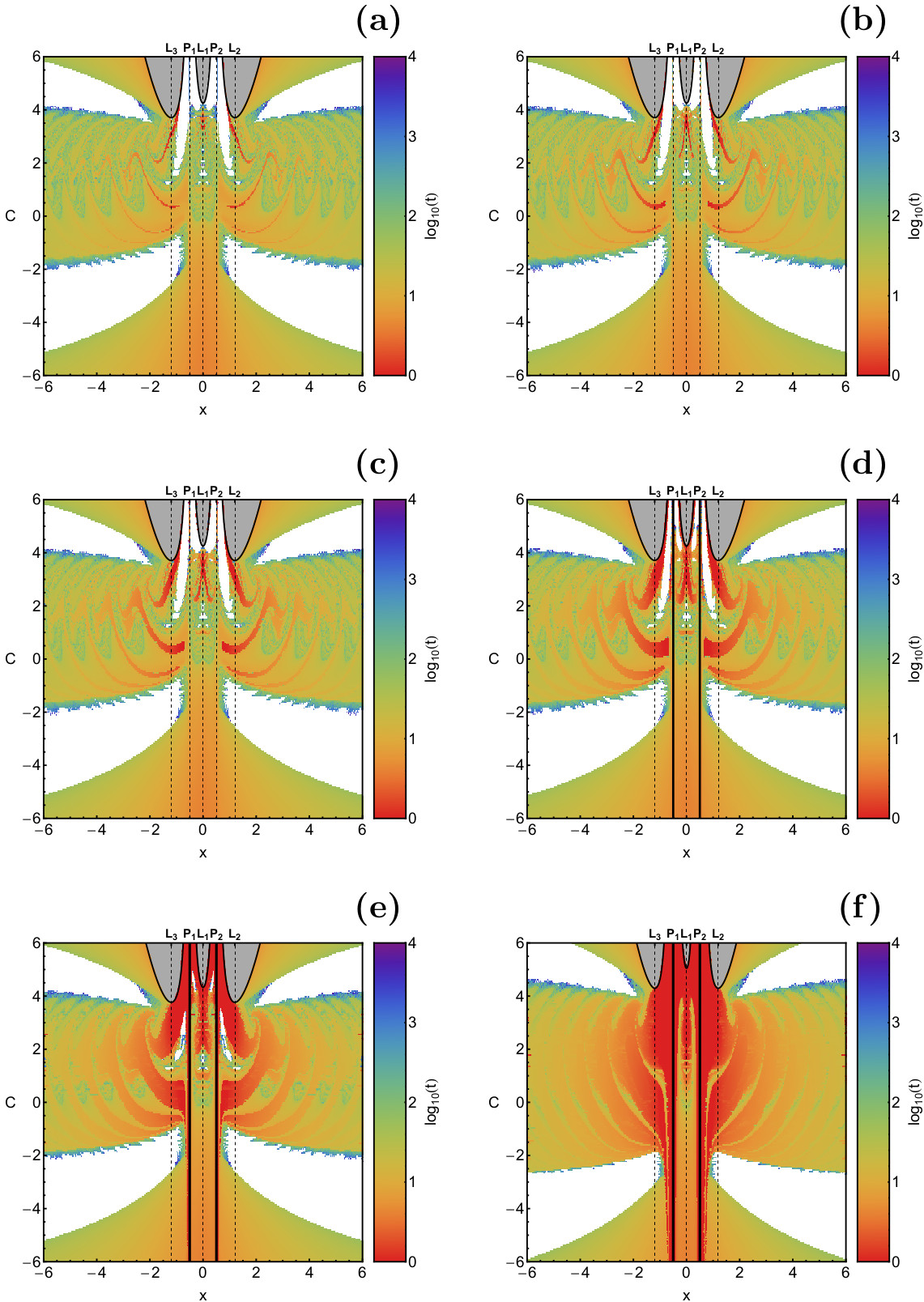}}
\caption{The corresponding distribution of the escape and collision time of the orbits for the panels of Fig. \ref{xCA}. (Color figure online.)}
\label{xCAt}
\end{figure*}

Our first case concerns the scenario where all the transport channels are closed. In Fig. \ref{HR1}(a-b) the CCDs reveal the orbital structure of the configuration $(x,z)$ space for two values of the Jacobi constant $C$. The black solid lines denote the zero velocity curves, while the inaccessible energetically forbidden regions are marked in gray. When $C = 5$, it is seen in panel (a) of Fig. \ref{HR1}, that inside the disks around the primaries the test particle can perform only two types of motion. More specifically, it can either move in bounded orbits around each primary, or collide with them. The SALI method indicates that the vast majority of the bounded orbits corresponds to non-escaping regular motion, while thin layers of trapped chaotic motion are also present, mainly in the vicinity of the basin\footnote{A local set of initial conditions of orbits which with the same final state is called a ``basin".} boundaries of the stability islands. Outside the energetically forbidden regions vast basins of escaping orbits exist, while there are also stability islands. Additional numerical calculations suggest that these regular orbits circulate around both primary bodies. In panel (b) of Fig. \ref{HR1} we observe that when $C = C_1$ there are several differences regarding the orbital structure of the $(x,z)$ plane, with respect to panel (a). In particular: (i) the amount of collision orbits inside the central regions has been increased; (ii) the presence of trapped chaotic orbits in the central regions is weaker; (iii) the stability islands in the exterior region (which correspond to regular motion around both primaries) have come closer to the energetically forbidden regions.

Because the two primary bodies have equal masses $(\mu = 1/2)$, as well as the same degree of oblateness $(A_1 = A_2)$, the dynamical system has several symmetries. Some of them are in fact displayed in the configuration $(x,z)$ plane, where as we can see in Fig. \ref{HR1}, the structures of the CCDs are symmetrical with respect to both the horizontal and the vertical axes. These symmetries are revealed due to the specific choice of the initial conditions of the orbits, explained in the previous section. Indeed, using the specific choice of the initial conditions the several types of basins display the expected symmetries. For example, the collision basins to two primaries are exactly the same (meaning that the amount of orbits which lead to collision with primary $P_1$ is exactly the same with the amount of orbits which collide with primary $P_2$). The same property applies also for the bounded basins, corresponding to the two primaries. As far as we know, this is the only choice of initial conditions for which the CCDs, on the $(x,z)$ plane, display such symmetries, regarding the several types of basins.

In the following Fig. \ref{HR1t}(a-b) we illustrate how the escape and collision time of orbits are distributed on the configuration $(x,z)$ space for the two values of the Jacobi constant discussed in Fig. \ref{HR1}(a-b). Light reddish colors correspond to fast escaping/collision orbits, dark blue/purple colors indicate large escape/collision time, while white color denote initial conditions of both regular non-escaping and trapped chaotic orbits. Note that the scale on the accompanying color bar is logarithmic. We see that some orbits collide with the primaries almost immediately. Looking more carefully at panel (b) of Fig. \ref{HR1t} we clearly observe that for a small portion of escaping orbits the corresponding escape times are at least two orders of magnitude larger than those of the vast majority of the escaping basins. These initial conditions of escaping orbits with high escape periods are mainly located in the vicinity of the boundaries between escaping and bounded basins.

\subsection{Energy case II $(C_2 \leq C < C_1)$}
\label{ss2}

The next case considers the second Hill's region configurations in which the test particle is allowed to move between the two primary bodies, inside the interior region, through the open neck around $L_1$. The orbital structure of the configuration $(x,z)$ space is unveiled in Fig. \ref{HR2}(a-b) through the CCDs. In panel (a), where $C = 4.22$, one can see that the orbital structure is very similar to that observed earlier in Fig. \ref{HR1}b for $C = C_1$. Around the saddle point $L_1$ there is a fractal\footnote{We would like to clarify that when we use the term ``fractal" for describing an area it means that this particular local region displays a fractal-like geometry. For more information the fractal dimension \cite{AVS01,AVS09} should be computed.} mixture of initial conditions which correspond either to collision or bounded motion. As the value of the Jacobi constant decreases, which means that the corresponding total orbital energy increases, there are two significant changes regarding the orbital content of the configuration $(x,z)$ plane. More precisely, in panel (b) of Fig. \ref{HR2}, which corresponds to $C = C_2$, it is seen that initial conditions of orbits that collide to primary 1 are present in the realm of primary 2 and vice versa. The other noticeable alteration concerns the geometry of the stability islands located in the exterior region.

The corresponding distribution of the escape and collision times of orbits on the configuration $(x,z)$ space is shown in Fig. \ref{HR2t}(a-b). We observe that the escaping orbits, with initial conditions in the exterior region, can be divided into two categories: (i) those, just outside the energetically forbidden regions, with relatively low escape periods ($t_{\rm esc} < 50$ time units), corresponding to dark blue colors and (ii) those with considerable higher values of escape rates ($t_{\rm esc} > 100$ time units) which correspond to milder intermediate colors.

\subsection{Energy case III $(C_4 \leq C < C_2)$}
\label{ss3}

Energy case III concerns the third Hill's region configurations where the test particle, with initial conditions in the
interior region, can escape from the system through the escape channels in the ZVSs, located near the Lagrange points $L_2$ and $L_3$. The following Fig. \ref{HR3}(a-d) shows the orbital structure of the configuration $(x,z)$ space, for values of the Jacobi constant. When $C = 3.7$ it is seen in panel (a) of Fig. \ref{HR3} that the pattern of the configuration $(x,z)$ plane is almost identical to that of panel (b) of Fig. \ref{HR2}. Even though the necks around the saddle points $L_2$ and $L_3$ are open there is no numerical indication of communication between the interior and the exterior realms. In particular, initial conditions of collision and escaping orbits are present only in the interior and in the exterior region, respectively. It seems that for such energy levels the two escape channels, around $L_2$ and $L_3$, are not wide enough so as to allow orbits to escape, even if this option is theoretically allowed. As the value of the Jacobi constant decreases the structure of the $(x,z)$ plane changes drastically (see panels (b-d) of Fig. \ref{HR3}).

The most important changes that occur are the following: (i) the channels around the saddle points $L_2$ and $L_3$ become more and more wide; (ii) the stability islands in the exterior region split off into four pieces; (iii) initial conditions of escaping orbits appear in the interior region, while at the same time initial conditions of collision motion populate the exterior realm; (iv) the initial conditions in the exterior region, corresponding to collision motion, seem to form waves which pass through the openings of the exterior stability islands. The formation of the wave-shaped structures is due to the specific choice of the initial conditions. Evidently, using other type of initial conditions (see e.g., \cite{BBS09b}) the shape of the collision basins will be different.

In Fig. \ref{HR3t}(a-d) we illustrate the corresponding distribution of the escape and collision time of orbits on the
configuration $(x,z)$ space. This type of plot allows us to monitor in more detail the evolution of the structure of the
central stability islands. We observe that when $C = C_4$ there exist only two stability islands in the central region,
corresponding to non-escaping orbits that circulate around each primary body.

\subsection{Energy case IV $(C < C_4)$}
\label{ss4}

The last energy case involves the scenario when $C < C_4$. Fig. \ref{HR4}(a-d) reveals the orbital structure of the configuration space through the CCDs. In panel (a) of Fig. \ref{HR4} it is seen that when $C = 2$ the vast majority of the configuration $(x,z)$ plane is covered by initial conditions of escaping orbits. Inside the vast escaping basin we can distinguish a plethora of smaller basins of initial conditions corresponding to collision motion. Furthermore, bounded motion is still possible. Indeed, stability islands of non-escaping regular orbits circulating around each and both primaries are present. According to panel (b) of Fig. \ref{HR4}, when $C = 0.5$ there are three main differences with respect to the previous case: (i) the energetically forbidden regions have disappeared, (ii) the amount of collision orbits has been reduced from about 24\% to 6\% and (iii) non-escaping regular motion around each primary body is no longer possible. With decreasing value of the Jacobi constant the portion of initial conditions which correspond to collision motion continues to reduce. Finally when $C = -2$ (see panel (d) of Fig. \ref{HR4}) collision motion is almost negligible, while the $(x,z)$ plane is dominated only by escaping and non-escaping regular orbits around both primaries. The corresponding distribution of the escape as well as the collision times of orbits on the configuration space is depicted in Fig. \ref{HR4t}(a-d). One can see similar outcomes with that presented earlier in the previous subsections.

\subsection{An overview analysis}
\label{over}

The CCDs in the $(x,z)$ space provide sufficient information for only a fixed value of the Jacobi constant. It would be very illuminating if we could scan a continuous spectrum of Jacobi constants $C$ rather than a few discrete values. In order to do this, we use new types of two-dimensional planes in which the value of the Jacobi constant is the ordinate, while the abscissa will be the $x$ or the $y$ coordinate of the orbits \cite{H69}. In particular, in the $(x,C)$ plane all orbits are launched from the $x$-axis, with $z = z_0$ and $y_0 = \dot{x_0} = \dot{z_0} = 0$, while the initial value of $\dot{y}$ is always derived through the Hamiltonian (\ref{ham}), in the form of Eq. (\ref{ini}). Similarly, in the $(y,C)$ plane the initial conditions are started form the $y$-axis, with $z = z_0$ and $x_0 = \dot{y_0} = \dot{z_0} = 0$, while the initial value of $\dot{x}$ is obtained as $\dot{x_0}(x_0 = 0, y_0, z_0;C) = \sqrt{2\Omega(x_0 = 0, y_0, z = z_0) - C}$.

In Fig. \ref{xyC}(a-f) we present the orbital structure of the $(x,C)$ and $(y,C)$ planes, for three values of $z_0$, when $C \in [-6,6]$. The black solid lines are the limiting curves which distinguish between regions of energetically allowed and forbidden motion and they are defined as follows
\begin{align}
f_1(x,C) = 2\Omega(x, y = 0, z = z_0) = C, \nonumber \\
f_2(y,C) = 2\Omega(x = 0, y, z = z_0) = C.
\label{zvcs}
\end{align}

In panels (a) and (b), where $z_0 = 0.01$ we observe that between the energetically forbidden regions, regular bounded basins corresponding to quasi-periodic orbits around one of the primaries are present. In the exterior region we have a strong presence of stability islands which correspond to ordered orbits that circulate around both primaries. Undoubtedly, the most interesting orbital structure exists between these stability islands, whose shape resembles a claw. Indeed, in these regions several basins of collision coexist with a large portion of basins of escape. Outside the bounded basins of the exterior region a escaping motion always dominates. The distribution of the corresponding escape and collision time of the orbits is depicted in Fig. \ref{xyCt}(a-f).

As we proceed to higher values of the initial coordinate $z_0$ the following changes, regarding the orbital structure of the $(x,C)$ and $(y,C)$ planes, take place: (i) the three energetically forbidden regions merge together, thus forming a unified forbidden region the area of which increases with increasing $z_0$ (ii) non-escaping regular motion around each primary is no longer possible, due to the unification of the energetically forbidden regions, (iii) the number of collision orbits, with initial conditions between the stability islands of the exterior region, is reduced, and (iv) the edges of the stability islands in the exterior region come closer and tend to merge near the boundaries of the energetically forbidden region.

In panels (a-f) of Fig. \ref{pxyC} we present the evolution of the rates of all types of orbits, as a function of $C$. Looking the diagrams of this figure we can draw the following conclusions:
\begin{enumerate}
  \item Escaping motion dominate (with percentages larger than 90\%) at very low $(C > 6)$, very high $(C < -5.5)$ and also at intermediate energy levels $(-1.5 < C < 3.5)$.
  \item Collision motion exist mainly when $-1.5 < C < 3.5$, while for all other values of the Jacobi constant the
      corresponding percentages tend to zero.
  \item Bounded regular motion exhibits two peaks at the exact energy levels where the lowest rates of the escaping orbits
      are observed.
\end{enumerate}

\section{Influence of the oblateness coefficient}
\label{obl}

So far we have established the influence of the Jacobi constant (energy level) as well as of the initial value of the $z$-coordinate on the nature of the orbits. In \cite{Z15a} and \cite{Z15c} we demonstrated how the oblateness coefficient affects the character of the orbits in the planar version of the Copenhagen problem. Therefore it would be of great interest to determine the influence of the same dynamical quantity on the orbital structure of the spatial version of the Copenhagen problem.

In order to obtain this information we classified sets of initial conditions or orbits on the $(x,C)$ plane, for six values of the oblateness coefficient, for fixed $z_0 = 0.01$. Our results are depicted in Fig. \ref{xCA}(a-f), while the corresponding distributions of the escape and collision time of the orbits are given in Fig. \ref{xCAt}(a-f). It is visible that as the primary bodies become more oblate several important changes occur regarding the orbital structure of the system. In particular:
\begin{enumerate}
  \item The portion of the initial conditions of orbits, which lead to collision to one of the primaries, significantly increases.
  \item The stability basins, corresponding to non-escaping regular motion around each primary, are reduced and for extremely high values of the oblateness ($A > 0.01$) they completely disappear. On the other hand, the stability islands, corresponding to regular orbits around both primaries, seem to be completely unaffected by the shift of the oblateness coefficient.
  \item The fractality of the phase space is considerably been reduced. Indeed, all the regions, containing fractal mixtures of different types of orbits, vanish due to the emergence of extended well-defined basins of collision.
\end{enumerate}

Strikingly enough similar behavior, regarding the influence of the oblateness coefficient on the character of the orbits, has also been reported in \cite{Z15a} and \cite{Z15c}. Therefore we may conclude that the oblateness of the primaries affects, in a very similar way, the nature of the orbits in both 2-dof and 3-dof versions of the circular restricted three-body problem.

Our numerical calculations indicate that the oblateness affects also the escape as well as the collision time of the orbits. Specifically, we observed that both the escape and the collision time of the orbits are reduced (at least for certain energy ranges) with increasing value of the oblateness coefficient.

As can be seen in Fig. \ref{pxyC}, the blue (collision with primary $P_1$) and the red (collision with primary $P_2$) curves lie directly on top of each other such that the blue is not visible. This is of course due to the symmetries of the problem ($\mu = 1/2$, and $A_1 = A_2$). As can be seen in Fig. \ref{xCA}, the phase space diagrams (i.e. the orbital structure) is indeed anti-symmetric regarding red and blue collision basins, as it should be for $A_1 = A_2$ and $m_1 = m_2 = 1/2$, given that the $(x,C)$ plane is displayed but the primaries are at positions $(x = - \mu, y = 0)$ and $(x = 1 - \mu, y = 0)$ in the corotating frame. Or in short, the primaries are located at $x = -1/2$ and $x = 1/2$, hence the observed anti-symmetry, regarding the $x$-position (for $A_1 = A_2$).

\begin{figure}[!t]
\centering
\resizebox{\hsize}{!}{\includegraphics{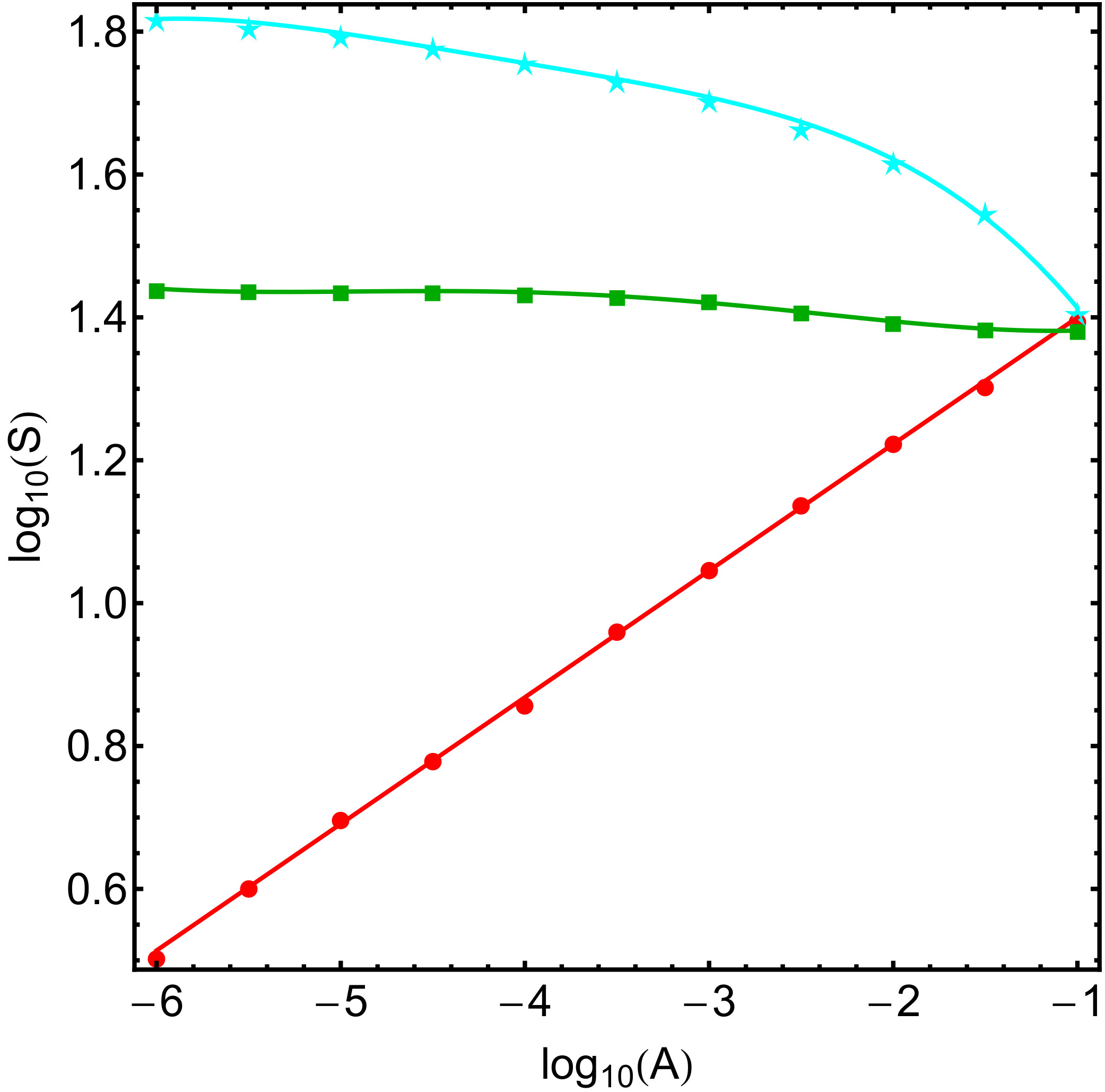}}
\caption{Evolution of the area $S$ of the different types (bounded, escape and collision) of basins on the $(x,C)$ plane, as a function of the oblateness coefficient $A$. The color code is as follows: escaping motion (cyan); bounded motion (green); collision motion to primary $P_1$ (blue); collision motion to primary $P_2$ (red). Note that because the amount of collision orbits to primary $P_1$ completely coincides with the amount of orbits that lead to collision with primary $P_2$ the corresponding percentages are the same and therefore the two curves on the diagram (blue and red) are overlapping each other. (Color figure online).}
\label{evol}
\end{figure}

Very similar outcomes can be extracted also from the $(y,C)$ planes. However, mainly for saving space, we do not present the corresponding diagrams. In contrast, it is illuminating to discuss the relation between the area of the different types of basins and the value of the oblateness coefficient $A$. Thus, in Fig. \ref{evol} we depict the evolution of the area $S$ of the different types of basins on the $(x,C)$ plane (see Fig. \ref{xCA}), as a function of $A$. In this diagram we do not include the parametric evolution of the trapped chaotic orbits because the corresponding percentages are extremely low. We observe that the area of the non-escaping regular orbits remains almost unperturbed and only for high values of $A$ displays a small reduction. Moreover, the area of the basins of escape is constantly reduced with the increase in the value of the oblateness coefficient. Finally, the most interesting behavior is that of the collision orbits, where it is seen that the area of the corresponding basins exhibits a linear increase in a log-log scale, i.e., a power law. At this point, it should be noted that in \cite{N04} (see Fig. 6) and \cite{N05} (see Fig. 11) a similar behavior was observed, regarding the area of the basins of collision and the corresponding radius of collision. When $\log_{10}(A) = -1$ collision orbits (to both primaries) occupy half of the $(x,C)$ plane, while non-escaping regular and escaping orbits share the rest half of the same plane.

In Fig. \ref{evol} we find a power law $S_{\text{collision}} = A^{\gamma}$, with $\gamma = 0.181$. In what follows we try to explain the power law behavior.

For the classical restricted three-body problem an approximation for the power law behavior of the area of the collision orbits in two-dimensional (2D) Poincar\'{e} surfaces of sections, $S_{\text{collision}} \sim R^\beta$, was derived, predicting the exponent $\beta = 1/2$ (see \cite{N04}). The RTBP can be roughly approximated by the Kepler problem when the test particle is close to one primary body, just before a collision occurs, or when the rotation of the primaries is neglected. Following the approach used in \cite{N04} the total orbital energy, in the inertial system, is given by $E_{\text{in}} = E + L$. In our notation we have $C = -2E$.

From Kepler's ellipse expression we obtain
\begin{equation}\label{eq:Kep1}
E = - \frac{1}{r_a + r_p}\pm \left(\frac{2r_p r_a}{r_p + r_a}\right)^{1/2},
\end{equation}
where $r_p$ denotes the perihelion and $r_a$ the aphelion distance. Solving Eq.\ (\ref{eq:Kep1}) for $r_a$ yields
\begin{equation}\label{eq:Kep2}
r_a(r_p) = - r_p - \frac{E + r_p^2}{E^2 - 2r_p} - \frac{\left(r_p^4 + 2E r_p^2 + 2r_p\right)^{1/2}}{E^2 - 2r_p}.
\end{equation}

A collision occurs when the test particle intersects the circle of radius $R$ around the Kepler singularity, $r_p \le R$. Hence, for $R\ll 1$, $S_{\text{collision}} \approx 2 \pi r_a(0) \left( r_a(0) - r_a(R)\right)$. Assuming $r_p \ll 1$,  Eq.\ (\ref{eq:Kep2}) shows approximately a square root behavior as a function of $r_p$, $r_a(r_p)\approx - \frac{1}{E} + \frac{(2 r_p)^{1/2}}{E^2}$. For the collision orbit area $S_{\text{collision}}$, this translates to
\begin{equation}\label{eq:Kep3}
S_{\text{collision}} \sim R^{\beta},
\end{equation}
with the exponent $\beta = 1/2$.

To link this result to the collision area as a function of the oblateness coefficient $A$, as presented in Fig. \ref{evol}, it is useful to recall the definition of the oblateness coefficient,
\begin{equation}
A = \frac{R_e^2 - R_p^2}{5R_d^2},
\label{obld}
\end{equation}
where $R_e$ is the equatorial radius and $R_p$ denotes the polar radius, while $R_d$ is the distance between the centers of the two primaries \cite{SSR76}. For a large equatorial oblateness, we find approximately that $A \sim R_e^2 := R^2$. Thus, using Eq. (\ref{eq:Kep3}) this provides a rough approximation for the collision area as a function of the oblateness coefficient, giving the power law $S_{\text{collision}} \sim R^{\beta} \sim  A^{\beta/2} = A^{\gamma}$, with $\gamma = 1/4$.

The power law fit of Fig. \ref{evol} gives $\gamma = 0.181$, which is a lower exponent that obtained from the rough approximation. This, however, may be due to the fact that the approximation does not account for the nonlinear dependencies between $A$ and the radii in Eq. (\ref{obld}), which may crucially determine the dynamics, collision orbits included. For the same reason, as it is clear from Fig. \ref{evol}, the areas of the three classes in Fig. \ref{evol} depend on the oblateness coefficient and do not add up to a constant ($S_{\text{collision}} + S_{\text{bounded}} + S_{\text{escape}} \neq const.$).

\section{Conclusions}
\label{conc}

In this paper we have studied the character of motion of a test particle in the three dimensional Copenhagen problem, where the two primary bodies are oblate spheroids. Our thorough and systematic numerical investigation allowed us to distinguish between bounded (with further classification between non-escaping regular and trapped chaotic orbits), unbounded and collision motion. We have studied the basins of escape and collision and related them to the corresponding escape and collision time of the orbits. Our analysis indicates that the three dimensional (3D) motion of the test particle in the system of two oblate spheroid primaries strongly depends on three main parameters: (i) the total orbital energy or equivalently the value of the Jacobi constant and (ii) on the initial value of the vertical $z$-coordinate of the orbits and (iii) the value of the oblateness coefficient.

To our knowledge this is the first time that the orbital dynamics, in the three dimensional Copenhagen problem with oblateness, is numerically investigated in a systematic manner, which represents our main contribution to existing literature.

The following list contains the most important conclusions of our numerical analysis.
\begin{enumerate}
  \item Escaping motion completely dominates at both very low as well as very high levels of the total orbital energy. Furthermore, escaping motion is the only type of motion which is possible for all values of the energy.
  \item For intermediate values of the energy we detected a very complicated mixture of all types of possible motion. In this regime all the types of basins exist and they display highly fractal basin boundaries.
  \item The vast majority of the initial conditions corresponding to non-escaping bounded motion corresponds to regular orbits. In fact there are two main types of regular motion: (i) motion only around one of the primary bodies and (ii) motion around both primaries. Our numerical computations showed that regular motion mainly corresponds to simple loop orbits (not shown here), while stability islands of secondary resonances\footnote{In this study, with the term secondary resonances we refer to all regular orbits with a shape more complicated than a simple loop.} are limited.
  \item Collision motion was found to be the only type of motion which is strongly related with the initial value of the $z$-coordinate of the orbits. In particular, the higher the $z_0$ of the orbits the lower the corresponding amount of detected collision orbits.
  \item Our calculations suggest that the fractality of the phase space depends on the initial value of the $z$-coordinate of the orbits. More precisely, we found fair numerical evidence that the phase space becomes less fractal as we proceed to higher levels of the $z$-coordinate.
  \item The oblateness coefficient $A$ was found to influence mostly the collision orbits. In particular, the amount of the collision orbits increases rapidly, with increasing value of the oblateness coefficient.
  \item The average escape and collision time of the orbits seems to be also connected with the oblateness coefficient. In fact, for specific energy regions, the escape and collision times were found to reduce with increasing value of $A$.
\end{enumerate}

Using an Intel$^{\circledR}$ Quad-Core\textsuperscript{TM} i7 2.4 GHz PC the required CPU time, for the classification of each grid of $1024 \times 1024$ initial conditions, was varying between 14 hours and 5 days, depending of course on the percentage of escaping and collision orbits. We would like to point out that the total computational cost may be reduced to 1/2 (or even to 1/4) if we take into account all the symmetries of the system. Nevertheless, in our computations we conducted the full run just for being sure that everything works fine, regarding the choice of the initial conditions and the symmetries of the system.

All the results as well as the conclusions of our numerical exploration are considered as a step towards the continuous task of revealing and understanding the great complexity of the orbital dynamics of the spatial circular RTBP with perturbations (e.g., oblateness). It is in out future plans to continue the numerical exploration of the system with three degrees of freedom (3-dof) by considering other sets of initial conditions of orbits (e.g., $\left(x_0,0,0,0,\dot{y_0},\dot{z_0}\right)$), so as to cover and unveil additional solutions of the system.

\section*{Acknowledgments}

Our warmest thanks go to two anonymous referees for the careful reading of the manuscript as well as for all the apt suggestions and comments which allowed us to improve both the quality and the clarity of the paper.

\begin{appendix}

\section{Terminology}
\label{apex}

\begin{figure}[!t]
\centering
\resizebox{\hsize}{!}{\includegraphics{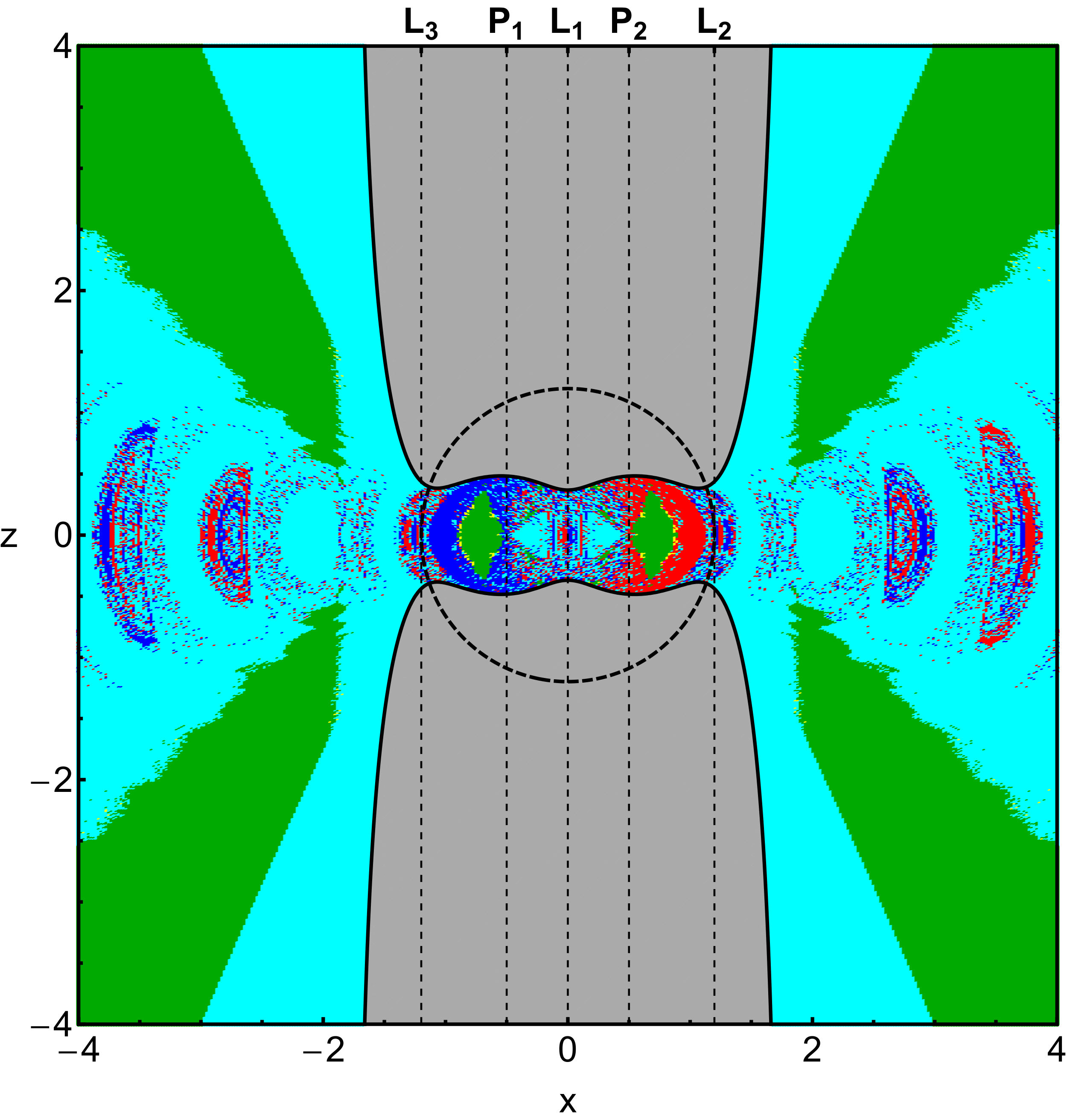}}
\caption{Color-coded diagram on the $(x,z)$ plane, when $C = 3.47$. The definition of all the parts of the diagram are described and explained in this Appendix. (Color figure online).}
\label{term}
\end{figure}

In this Appendix we will provide the explanation as well as the definitions of some of the specialized terminology used in this paper. Fig. \ref{term} will be used as a test figure in order to correlate the terms with the corresponding parts of the color-coded diagrams.

\begin{itemize}
  \item \textbf{Energetically allowed regions:} For every value of the Jacobi constant $C_0$ the areas of the phase space where $2 \Omega(x,y,z) > C_0$ are called energetically allowed regions. In these regions the test particle is permitted to move, according to the equations of motion. On the other hand, the regions of the phase space where $2 \Omega(x,y,z) < C_0$ are called energetically forbidden regions and the test particle is not allowed to enter them. The energetically forbidden regions are always shown in gray in the color-coded diagrams.
  \item \textbf{Interior realm:} The subregion of the energetically allowed areas between the two saddle points (that is for $x(L_3) < x < x(L_2)$) defines the interior realm or the central region of the phase space. In Fig. \ref{term} the interior realm is defined as the area inside the black, dashed circle. All the energetically allowed areas outside this circle define the so-called exterior region or exterior realms.
  \item \textbf{Transport channels:} In Fig. \ref{term} it is seen that there are two disjoint parts of energetically forbidden regions. These parts enclose the interior region of the phase space. Moreover, one can see, that near the equilibrium points $L_2$ and $L_3$ the energetically forbidden regions form two channels through which orbits from the interior region can enter the exterior region or vice versa. These channels, located in the vicinity of the saddle points, are the transport channels of the dynamical system.
  \item \textbf{Stability islands:} The green regions on the color-coded diagrams are formed by a collection of initial conditions corresponding to non-escaping regular orbits. These regions are called stability islands or bounded basins. In this dynamical system there are two main types of stability islands or bounded basins: (i) those which are located in the interior region and correspond to regular orbits that circulate around only one primary and (ii) those which are located in the exterior region and are composed of initial conditions of orbits that circulate around both primary bodies.
  \item \textbf{Escaping and collision basins}: We observe in Fig. \ref{term} that a large portion of the $(x,z)$ plane is covered by initial conditions of escaping orbits which form the so-called basins of escape (cyan regions). In the same vein, the initial conditions of orbits which lead to collision with one of the primaries form the so-called basins of collision (blue and red regions). In some areas of the phase space the escape and collision basins are clearly distinguishable, while there are also regions where the initial conditions create a highly complicated fractal mixture of escaping and collision motion.
\end{itemize}

\end{appendix}

\footnotesize


\end{document}